\documentclass{KapProc}
\normallatexbib

\usepackage{epsfig} 

\begin{document}

\articletitle{From the Fermi liquid towards the Wigner solid 
in two dimensions}

\author{Jean-Louis Pichard $^a$, Giuliano Benenti, 
Georgios Katomeris \\
Franck Selva and Xavier Waintal}

\affil{Service de Physique de l'Etat Condens\'e, 
CEA-Saclay, 91191 Gif sur Yvette cedex, France}
\email{$^a$ Pichard@drecam.saclay.cea.fr}

\vskip1cm
\noindent
{\bf To appear in:} {\it Exotic States in 
Quantum Nanostructures}\, ed.\ by S.~Sarkar, Kluwer, Dordrecht.

\begin{abstract}

The quantum-classical crossover from the Fermi liquid 
towards the Wigner solid is numerically revisited, considering 
small square lattice models where electrons interact via a 
Coulomb $U/r$ potential. We review a series of exact numerical 
results obtained in the presence of weak site disorder for fully 
polarized electrons (spinless fermions) and when the spin degrees 
of freedom are included. A novel intermediate regime between 
the Fermi system of weakly interacting localized particles and 
the correlated Wigner solid is obtained. A detailed analysis of 
the non disordered case shows that the intermediate ground state 
is a solid entangled with an excited liquid. For 
electrons in two dimensions, this raises  the question of the 
existence of an unnoticed intermediate liquid-solid phase. 
Using the Coulomb energy to kinetic energy ratio 
$r_s \propto U \propto n_s^{-1/2}$, 
we discuss certain analogies between the numerical results obtained 
as a function of $U$ for a few particles and the low temperature 
behaviors obtained as a function of the carrier 
density $n_s$  in two dimensional electron gases. Notably, 
the new ``exotic state of matter'' numerically observed at 
low energies in small clusters occurs at the same intermediate 
ratios $r_s$ than the unexpected low temperature metallic behavior 
characterizing dilute electron gases. 
The finite size effects in the limit of strong disorder 
are eventually studied in the last section, providing 
two numerical evidences that the weak coupling Fermi limit 
is delimited by a second order quantum phase transition when 
one increases $U$.

\end{abstract}

\newpage


\vskip1cm

   ``The very simplest form of the theory of the energy bands in metals 
gave for many problems such accurate explanations of often very intricate 
properties of metals and alloys that it may well appear superfluous to 
consider extensions of the simple form of the theory''. Those words written 
by Wigner \cite{wigner} in 1938 come again as an objection against the need to 
develop a more rigourous theory, since the Fermi liquid theory 
(FLT) was improved by Landau \cite{landau} in the sense of a perturbation 
theory based on renormalized single-particle excitations and adapted to 
include the effects of elastic scattering by the impurities 
\cite{altshuler-aronov}. The need to go outside conventional 
FLT for explaining  the unexpected two dimensional metallic 
phase \cite{aks} discovered by Kravchenko and Pudalov is a subject of 
controversy. On one hand, certain characteristic FLT behaviors \cite{simmons}
seem to remain in the vicinity of the metal-insulator transition 
(MIT), suggesting that an ``apparent'' metallic behavior could be 
the consequence of ``classical'' effects (interband scattering \cite{sivan}, 
temperature dependent screening \cite{gennser},  temperature dependent 
scattering \cite{altshuler-maslov,savchenko} or classical 
percolation \cite{meir}). 
On the other hand, the observation of an unexpected MIT is 
first the result (see for instance Refs. \cite{mills,batllog}) 
of new possibilities of studying controlled many body systems 
which are closer to the strong coupling limit 
than the previously studied systems in two dimensions. 
This gave us the motivation to numerically revisit the classic 
problem of the crossover from the weak coupling Fermi limit 
towards the strong coupling Wigner limit for electrons in two 
dimensions. In this chapter, we review our main numerical results. 
The interest of the information given from exact diagonalization 
of small systems can be questionned, 
but it may have the merit to raise questions which may be relevant for 
explaining the behaviors observed around the ``two dimensional MIT''.

\section{weak and strong coupling limits}

  A convenient measure of the electron gas density $n_s$ is the 
dimensionless parameter $r_s$ 
\begin{equation}
r_s=\frac{1}{\sqrt{\pi n_s} a^*_B}
\end{equation}
defined as the radius $1/{\sqrt {\pi n_s}}$ of the unit disk divided by the 
Bohr radius $a_B=\hbar^2/(m e^2)$. The unit disk encloses an 
area equal to the area per electron of the gas. For a real two dimensional 
electron gas (2DEG) created in a field effect device, one uses an effective 
Bohr radius $a^*_B=\hbar^2 \epsilon/(m^* e^2)$ which includes the dielectric 
constant $\epsilon$ of the medium in which the 2DEG is created and 
the effective mass $m^*$ of the carriers. Wigner was the first to 
consider the dilute limit where $r_s$ becomes large,  
the Coulomb interactions dominate the kinetic energy
in determining the wave function and the electrons tend to arrange 
themselves in a regular lattice. It may be argued that a lattice 
configuration is not consistent with the translational symmetry 
characterizing the 2DEG Hamiltonian in the absence of a random substrate. 
This objection can be removed by forming a new wave function which is 
a linear combination of all translations of the original lattice: the 
resultant wave function will have a uniform electronic charge density, 
as symmetry demands, with an unchanged energy. Originally, Wigner assumed 
a bcc electron lattice. It was later shown \cite{maradudin} 
that the hexagonal lattice has a lowest electrostatic energy in two 
dimensions. 
 
 As explained in Refs. \cite{pines,carr,ceperley}, when one 
considers the Hamiltonian of $N$ electrons in two dimensions, 
\begin{equation}
{H} = \frac{1}{r_s^2} \sum_{i}^N \nabla_{i}^2 + \frac{2}{r_s} 
\sum_{i,j\atop i\neq j}^N \frac{1}{|r_i-r_{j}|}
\end{equation}
where the lengths are given in units of $1/\sqrt{\pi n_s}$, 
the question has been from the early days to obtain the 
asymptotic behaviors of the ground state 
energy $E_0$ around the weak coupling limit ($r_s<<1$): 
\begin{equation}
E_0 = \frac{h_0}{r_s^2}+\frac{h_1}{r_s}+0(\ln r_s)
\end{equation}  
and around the strong coupling limit ($r_s>>1$): 
\begin{equation}
E_0=\frac{f_0}{r_s}+\frac{f_1}{r_s^{3/2}}+\frac{f_2}{r_s^2}+0(r_s^{-5/2}),  
\end{equation}  
to calculate the coefficients $h$ and $f$, and to discuss the expected 
range of validity for those asymptotic expansions.  Then, one can try 
to numerically determine the value of $r_s$ where the weak coupling 
energy exceeds the large coupling energy. This can be done 
at the price of certain 
approximations which are controlled in the two limits and remain  
more uncertain in the middle, and after extrapolating finite size 
studies towards the thermodynamic limit. 
 
 The most advanced works in this field are Quantum Monte Carlo studies 
\cite{ceperley,tanatar-ceperley} of 
systems involving many electrons (typically more than $10^2$) and assuming 
two trial wave functions $\Psi_T(R)$ adapted to describe either the weak 
coupling limit, or the strong coupling limit. $\Psi_T(R)$ are of the 
Slater-Jastrow form 
\begin{equation} 
\Psi_T(R)=D(R) \exp (-\sum_{i<j}^N u(|r_i-r_j|)),
\end{equation} 
where $D(R)$ is a Slater determinant of extended plane waves for weak 
coupling, of localized single-particle orbitals for strong coupling. 
The liquid and crystal pseudopotentials $u(r)$ are repulsive and include 
in an approximate way the effects of electronic correlations. Then a simple 
variational approach, or a more involved fixed node Green's-function 
approach, are used to obtain the energies $E_0(r_s)$ dictated by the chosen 
$\Psi_T(R)$ or by its nodal structure. Both the variational energies and the 
fixed node energies give an upper bound to the exact energy. Comparing for 
intermediate $r_s$ the energies given by the $\Psi_T(R)$ adapted to describe 
the liquid and the crystal, one concludes \cite{tanatar-ceperley} that there 
is a first order quantum liquid-solid transition at $r_s \approx 37$, with a 
possible division \cite{ceperley,senatore} of the liquid phase into a 
non polarized liquid at small $r_s$ and a polarized liquid for larger $r_s$. 
However, the nature of the quantum mechanism of melting is still debated, 
and the possibility of a continuous transition has been very recently 
proposed\cite{candido}. The solid is assumed to be a frustrated 
antiferromagnet\cite{bernu} before 
becoming ferromagnetic \cite{roger} 
at very large $r_s$. The same Monte Carlo method 
has been used\cite{chui-tanatar} in the presence of impurities. The 
conclusion was that disorder can stabilize the solid to weaker values 
of $r_s$. 

 Andreev and Lifshitz have discussed \cite{AL} in 1969 the possibility 
to have a more complex intermediate state between the solid and the liquid, 
which should be neither a solid nor a liquid. Two kinds of motion should be  
possible in it; one possesses the properties of motion in an elastic 
solid, the second possesses the properties of motion in a liquid. This 
idea comes from a theory of defects in quantum solids. 
The nature of the relevant defects is a complicated issue. Let us 
give a possible example: a vacancy yielded by one electron hopping from 
the Wigner lattice towards some interstitial site. In a classical 
solid, this defect has a certain electrostatic cost and remains localized. 
In a quantum solid, we have in addition the tunneling effect, and if 
this defect can be created, it will be delocalized since the system 
is invariant under translation. Therefore, when $r_s$ decreases 
from the strong coupling limit, the increasing band width of the zero point 
defects of this type may exceed their decreasing electrostatic 
energy cost, leading to two possibilities for intermediate $r_s$:  
Either the total melting of the solid to directly give a liquid, as 
implicitely assumed for instance in Refs. \cite{candido}, or a quantum 
floppy solid coexisting with a liquid of delocalized defects, 
as conjectured by Andreev and Lifshitz. A phenomenological FLT theory 
\`a la Landau of such gapless ``delocalized excitations'' of a floppy 
quantum solid has been later proposed in Ref. \cite{DKL}. 
The discussion of this second possibility is one of the central points 
of this chapter. 

\section{Dilute 2DEG in field effect devices}
	
 A pure 2DEG can be realized by trapping electrons on the surface of 
 liquid helium, but it is difficult to reach a sufficient density 
 to study the quantum regime. Another possible realization is  
 given by the new classes of superconducting cuprates where the 
 electronic motion is essentially two dimensional. The charge density 
 can be varied by chemical doping, and a complex phase diagram is 
 obtained, with insulating, superconducting and metallic behaviors. 

\begin{figure}[ht]  
\vskip.2in
\centerline{\epsfxsize=12cm\epsffile{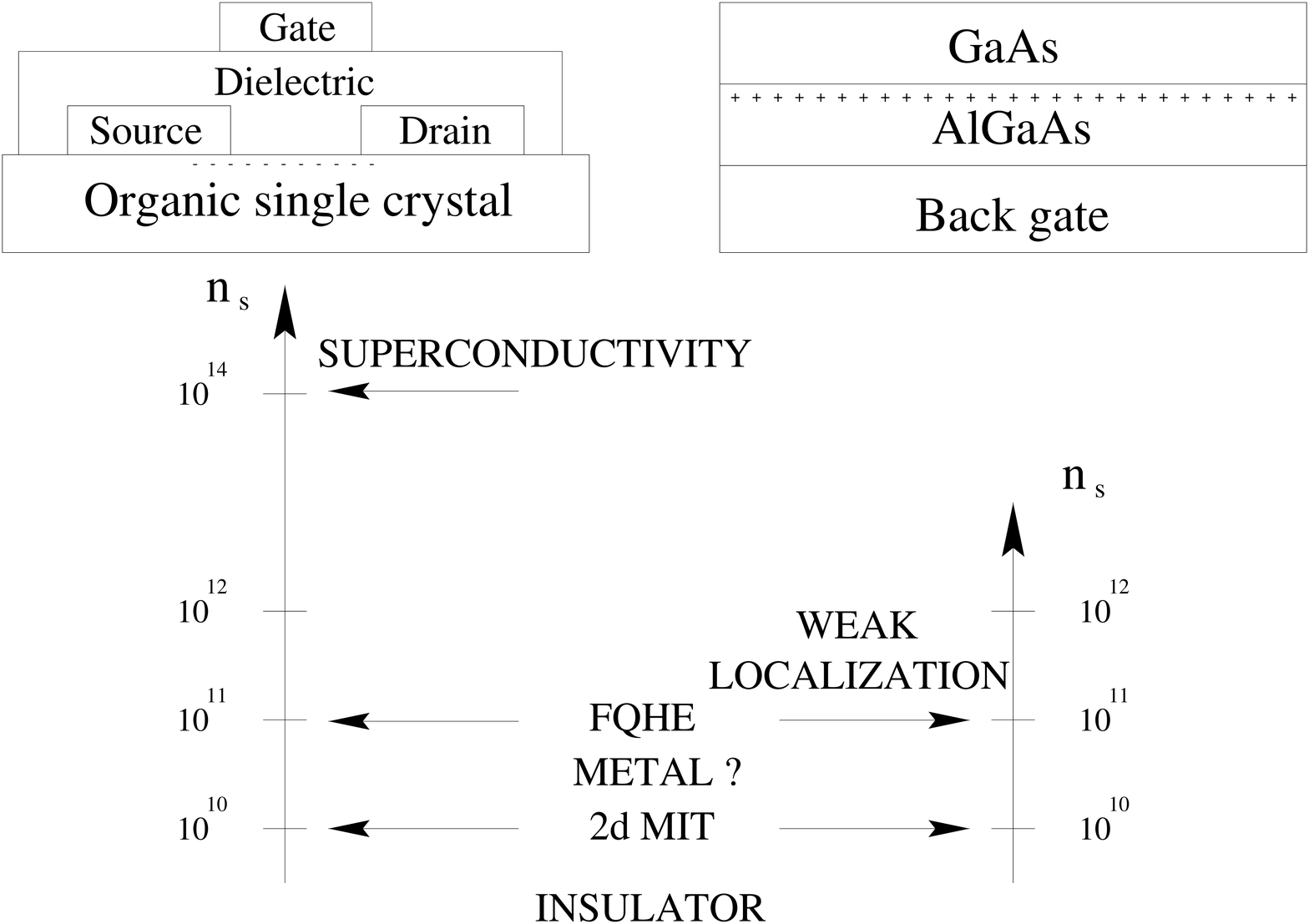}} 
\caption{Schematic picture of an organic field effect transistor 
(upper left) and of a GaAs heterostructure (upper right) where the 
location of the 2DEG or 2DHG (symbol plus for the holes) is indicated. 
The characteristic low temperature behaviors are summarized below 
as a function of the (typical) carrier density $n_s$ ($cm^{-2}$): 
2dMIT, $2d$ metal, FQHE for a sufficient magnetic field, weak 
localization correction to the Boltzmann conductivity, superconductivity 
(see Ref.\cite{batllog-supra}). The densities give a typical order of 
magnitude, the observed behaviors depending also on the effective mass 
of the carriers.}
\label{fig1}
\end{figure} 

 Eventually, one can create a two dimensional gas of 
 charges (electrons or holes) at the interface between two doped 
 semiconductors (GaAs-AlGaAs heterostructures), between a semi-conductor 
 and an insulator (Si-Mosfet), or very recently \cite{batllog} between 
 an organic crystal (pentacene, tetracene and anthracene) and an 
 insulator. The carrier density can be varied by a gate from a very 
 dilute limit towards larger densities. In Fig. \ref{fig1}, 
 we mention the remarkable phenomena observed in organic (left) or 
 doped semi-conductor (right) field effect devices. A clean interface 
 may give a high carrier mobility, may allow the observation 
 \cite{batllog-fqhe} of the fractional quantized Hall effect (FQHE) 
 and may give a measurable conductivity in a very dilute limit  
 (typically $n_s \approx 10^9-10^{11}$ carriers per cm$^2$) for GaAs 
 heterostructures and organic devices. If the effective mass of 
 the carriers is large enough (a condition which is not satisfied 
 by the electrons in Ga-As heterostructures) the effective factor 
 $r_s$ can be in the vicinity of the values where the Fermi-Wigner 
 crossover is expected. One of the surprises in those high quality 
 field effect transistors has been the observation of a metallic 
 low temperature behavior \cite{aks,batllog,Batllog} in a certain 
 intermediate range of carrier densities, where a large perpendicular 
 magnetic field yields FQHE or magnetically induced Wigner 
 crystals \cite{williams}.  

\section{Metal-insulator transition}

 A recent review of the 2D-MIT can be found in Ref. \cite{aks} with an 
extended list of references. We summarize by a few sketches some of the 
behaviors which have been observed, and which are useful for 
discussing our numerical results. The main surprise was given 
by the temperature dependence of the 2DEG resistivity $\rho(T)$ 
around a low critical density $n_{c1}$. 
\begin{figure}[ht]  
\vskip.2in
\centerline{\epsfxsize=12cm\epsffile{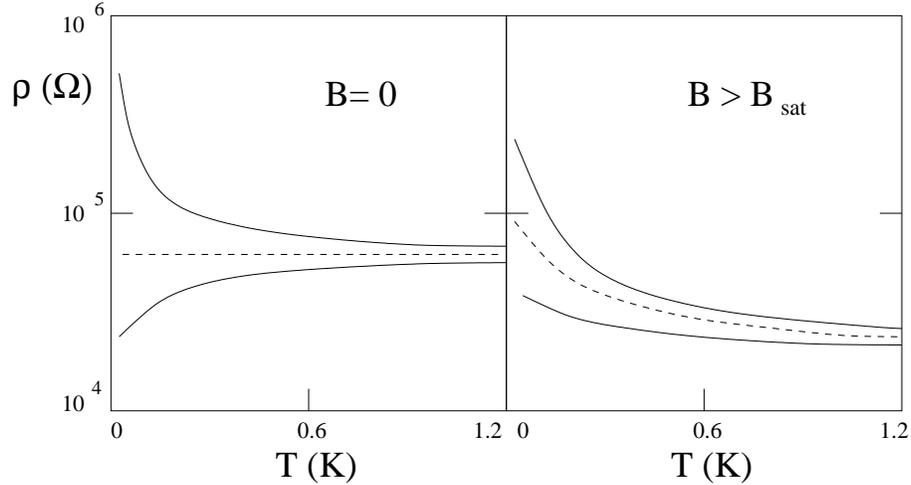}} 
\caption{Resistivity versus temperature without (left) and with (right) 
a large parallel magnetic field $B>B_{sat}$. $n_s < n_{c1}$ (upper curves), 
$n = n_{c1}$ (middle curves) and $n > n_{c1}$ (lower curves).  
}
\label{fig2}
\end{figure} 
As sketched in Fig. \ref{fig1} (see Fig.1 of Ref. \cite{shashkin1}), 
$\rho(T)$ decreases as a function of $T$ when $n_s < n_{c1}$, 
becomes temperature independent at $n_s=n_{c1}$ and increases 
when $n_s> n_{c1}$. A decay is the expected behavior for an insulator, 
while an increase usually characterizes a metal. 
These behaviors occur \cite{kk} in a low temperature range 
$35 mk < T \leq T_F$, where $T_F \approx 0.8 - 5 K$ are the typical 
Fermi temperatures of those dilute 2DEGs. The temperature increase of 
$\rho(T)$ can be large for a 2DEG created in a Si-Mosfet 
(typically one order of magnitude), but remains weak  
in a 2DHG created in a GaAs heterostructure.  For the densities $n_s$ 
where $\rho(T)$ has a metallic behavior, a parallel field $B$ induces 
a large positive magnetoresistance which saturates above a certain 
field $B_{sat}$, as sketched in Fig.\ref{fig5} (see Fig. 1 of Ref. 
\cite{vitkalov1} and Fig. 3 of Ref. \cite{vitkalov2}).  
From small angle Shubnikov-de Haas measurements done in a Si-mosfet, 
it was concluded in Ref. \cite{vitkalov1} that $B_{sat}$ signals also 
the onset of full spin polarization. Close to the MIT, $B_{sat}$ is 
very small and increases as $n_s-n_{c1}$ above $n_{c1}$ \cite{vitkalov2}. 
This corresponds to the intermediate values of $r_s$ 
(typically $3 <r_s < 10$) where the metallic behavior is observed.
\begin{figure}[ht]  
\vskip.2in
\centerline{\epsfxsize=12cm\epsffile{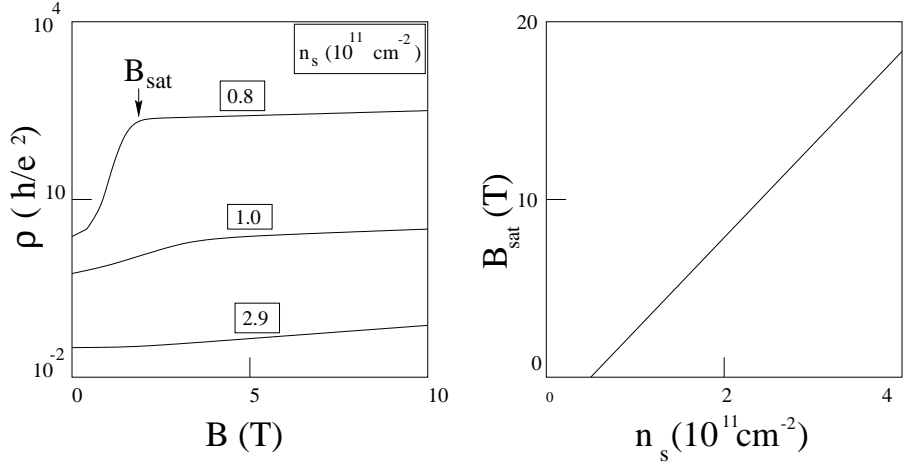}} 
\caption{Resistivity versus parallel magnetic field (left) and 
saturation field $B_{sat}$ as a function of the carrier density (right).}
\label{fig5}
\end{figure} 
When $B>B_{sat}$, the metallic increase of $\rho(T)$ disappears, 
but the $I-V$ characteristics sketched in Fig. \ref{fig3} 
(Fig 2 of Ref. \cite{shashkin1}) indicates  
the existence of a critical density $n_{c2}$ below which a non linearity 
is observed and above which it disappears. 
The density $n_{c1}$ and $n_{c2}$ are close to each others, if not 
identical when $B=0$. The dependence of the characteristic $n_{c2}$ 
as a function of a parallel magnetic field $B$ is sketched in 
Fig. \ref{fig3}. (Fig. 4 of Ref. \cite{shashkin1}). 

\begin{figure}[ht]  
\vskip.2in
\centerline{\epsfxsize=12cm\epsffile{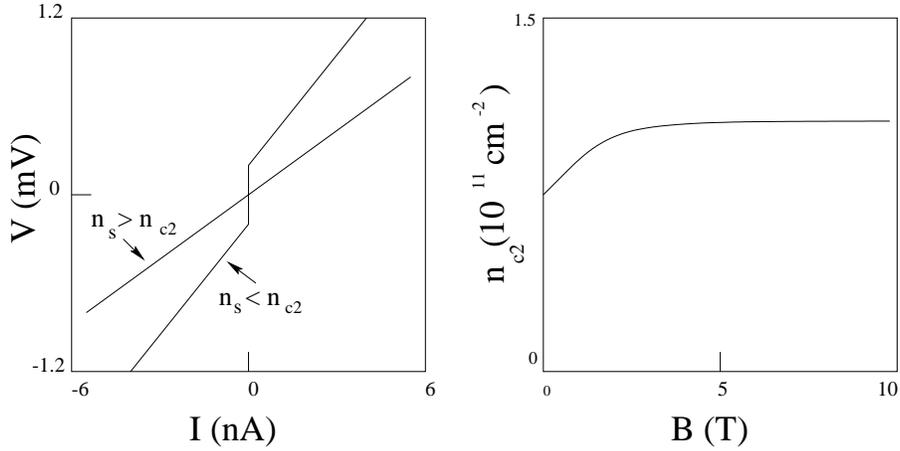}} 
\caption{Current-voltage non linear characteristics of the 
insulating phase which are suppressed above the MIT (left). 
Critical density $n_{c2}$ above which the non linearity of the 
I-V characteristics disappears as a function of the parallel 
magnetic field.}
\label{fig3}
\end{figure} 

\begin{figure}[ht]  
\vskip.2in
\centerline{\epsfxsize=7cm\epsffile{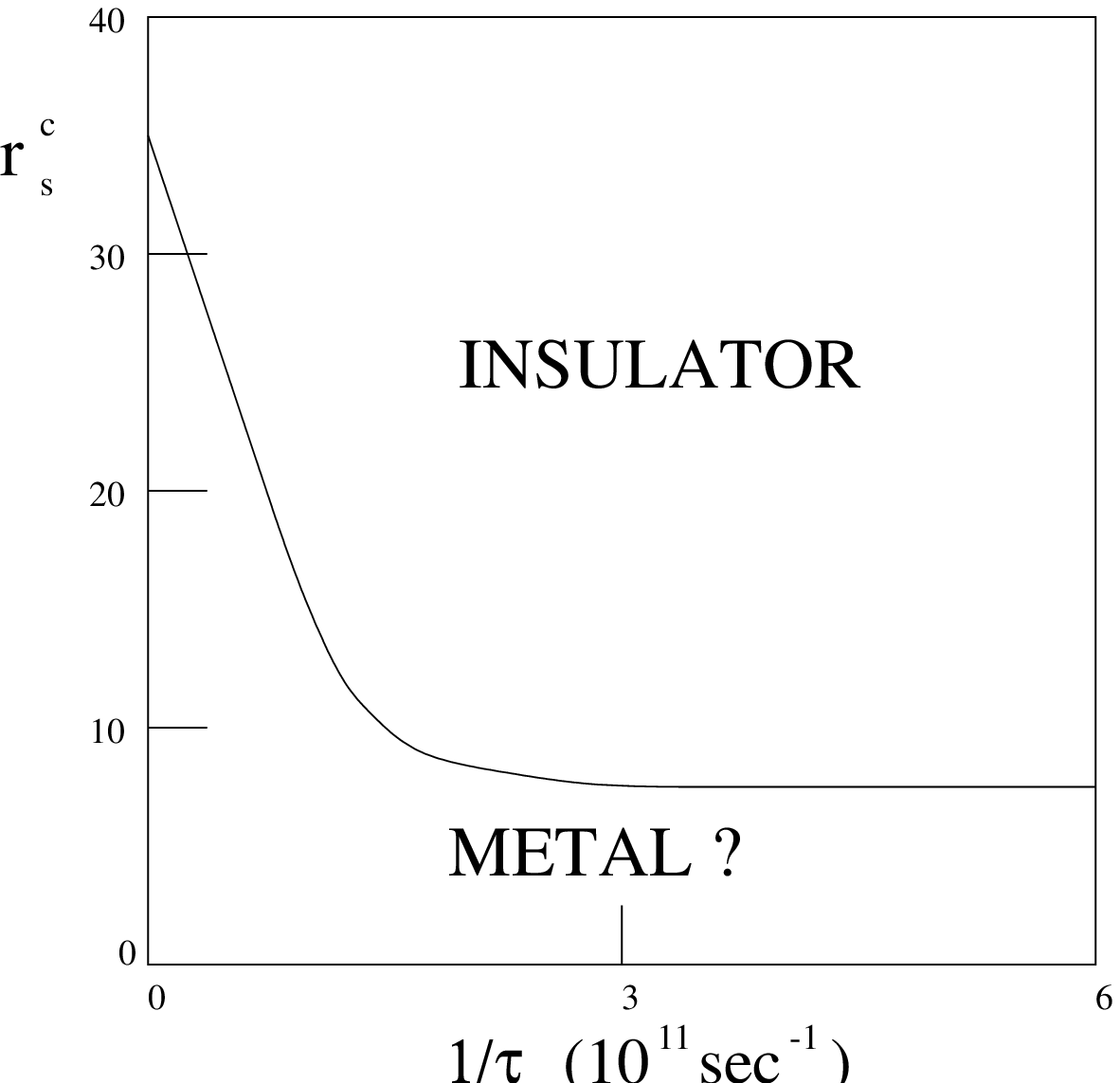}} 
\caption{Critical factor $r_s^c$ at which the MIT is observed as a 
function of the inverse elastic scattering time.}
\label{fig4} 
\end{figure}

The critical density $n_{c1}$ does not give a unique critical value 
for the factor $r_s$. Impurity scattering plays a role. For clean 
systems, one needs to have a much larger factor $r_s$ than in a dirty 
system, as sketched in Fig. \ref{fig4}. 
(see inset of Fig. 1 in Ref. \cite{yoon}).
In an (undoped) organic field effect transistor, the 2DEG is less 
scattered by impurities and the MIT is seen\cite{Batllog} at an even 
smaller density ($r_s \approx 50$).  

\begin{figure}[ht]  
\vskip.2in
\centerline{\epsfxsize=7cm\epsffile{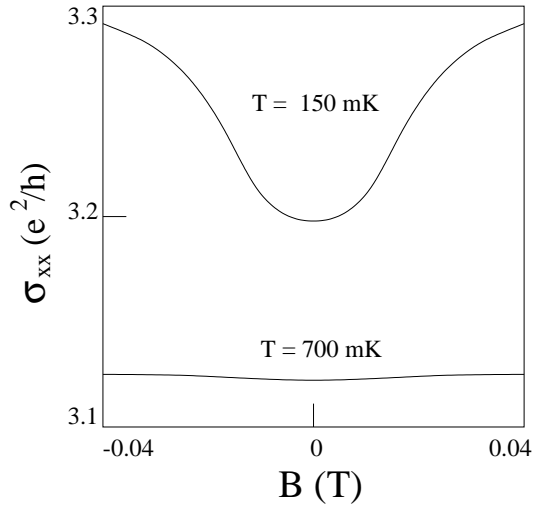}} 
\caption{Usual weak localization behavior of the conductance 
induced by a perpendicular magnetic field in a 2DHG created in 
Ga-As heterostructure.}
\label{fig6}
\end{figure}

The surprise caused by this unexpected metallic behavior was mainly 
due to its discrepancy with the scaling theory of localization, which 
does not take into account electron-electron interactions. When 
the conductance $g$ is 
larger than the conductance quantum $e^2/h$, a weak disorder perturbative 
expansion gives for the average conductance a universal logaritmic 
correction to the Drude conductivity which defavors transport, unless 
there is a sufficient spin orbit scattering. In the weak disorder limit, 
one can also take into account the interaction when $r_s < 1$ and one 
obtains additional 
corrections which reduce transport in a similar way. However, the 
extrapolation of the small $r_s$ interaction dependent correction 
to larger $r_s$ suggests a possible change of 
the sign of the corrections due to the interactions, indicating the 
possibility of a metallic phase in two dimensions, as mentioned by 
Finkelshtein \cite{finkel}. Computer calculations without interaction 
and transport measurements at not too low densities made in the eighties 
have confirmed the absence of metallic behavior in two dimensions. The 
difference between the recent experiments giving a MIT and the former 
experiments confirming the absence of metallic behavior seems to be the 
quality of the interfaces at which the 2DEG is created. This 
feature makes possible to 
have a measurable conductivity at much lower carrier densities than 
previously. An intermediate range of density, where the factor $r_s$ is too 
large (too small) to allow expansion in powers of $r_s$ ($1/r_s$) 
and a weak elastic scattering seems to be necessary for observing 
the metallic behavior. This hypothesis was supported by Ref. \cite{hamilton} 
where a study of a 2DHG in a Ga-As heterostructure gives a range 
$r_s^F < r_s < r_s^W$ for having a weak metallic behavior 
in a disordered sample. When $r_s<r_s^F$, one would have weakly interacting 
quasi-particles dominated by Anderson localization when the temperature 
$T \rightarrow 0$. When $r_s > r_s^W$, one would have a highly correlated 
set of charges. Between $r_s^F \approx 6$ and $r_s^W \approx 9$ in the 
studied sample, a problematic small metallic behavior is observed between 
two insulating behaviors of different nature (Anderson insulator for large 
densities, pinned Wigner solid for low densities).  The re-entrant MIT at 
low $r_s$ is not easy to observe, if it exists, since the localization 
length of a clean device can be very large, and the observation of a 
possible re-entrant insulating behavior at high densities can require very low 
temperatures. This is why many observations of a MIT have been reported 
for $r_s  \approx 10$, while very few experiments give a possible 
re-entrant MIT at $r_s \approx 3$. Moreover, more recent works  
\cite{simmons,gennser,savchenko}) put doubts about 
the reality of this intermediate metallic behavior when the temperature 
goes to zero, since the effect of a weak perpendicular magnetic 
field can be described by usual weak localization theories, even for 
values of $r_s$ as large as $15$ (see Fig. \ref{fig6} taken from Ref. 
\cite{simmons}).  
The hypothesis of a certain temperature dependent screening was suggested 
for explaining the anomalous temperature dependence, and it was proposed  
that usual quantum interferences should drive at possibly very low 
temperatures the system to the formerly expected insulating behavior. 
Recent measures performed down to $5 mK$ do not confirm\cite{bell-labs} 
this hypothesis. The estimate of the phase breaking length $L_{\phi}$ 
which is traditionally done for estimating the low temperature dependence 
of the resistance from a zero temperature theory leads again to the 
famous problem of the saturation \cite{webb} of $L_{\phi}$ when 
$T \rightarrow 0$, problem leading also to many possible and controversial 
explanations. 
 
 In summary, a significant metallic behavior can be seen using a 2DEG created 
in a Si-mosfet, while a weaker one occurs in a 2DHG created 
in a Si-Ge quantum well or a Ga-As hetrostructure. Nevertheless, in the 
latter system, the study of the compressibility gives complementary signatures 
\cite{jiang,yacoby1,yacoby2}  of a possible quantum phase transition. Local  
compressibility measurements show that the system is more homogenous in the 
intermediate metallic phase than in the low density insulating phase. 
Very recently, the possibility 
that the MIT would be accompanied by a magnetic transition has been 
suggested \cite{vitkalov2,raznikov}.     

\section{Lattice model}

 The previous experimental observations lead us to numerically revisit 
the Fermi-Wigner crossover using a two dimensional model describing 
$N$ particles on $L \times L$ square lattice with periodic boundary 
conditions (BCs), i.e. 
with a torus topology. The most general Hamiltonian ${\cal H}$ of the 
lattice model we will focus on reads, 
\begin{eqnarray}
{\cal H}&=& \sum_{i,\sigma} (-t \sum_{i'} c^{\dagger}_{i',\sigma}
c_{i,\sigma} + v_i n_{i,\sigma}) 
\nonumber \\
& &+\frac{U}{2} \sum_{i,i'\atop i\neq i'} 
\frac{n_{i,\sigma} n_{i',\sigma'}}
{|i-i'|}+ 2U 
\sum_{i} n_{i,\uparrow}n_{i,\downarrow},
\end{eqnarray}
where the operators $c_{i,\sigma}$ ($c^{\dagger}_{i,\sigma}$) destroy 
(create) an electron of spin $\sigma$ at the site $i$ and 
$n_{i,\sigma}=c^{\dagger}_{i,\sigma}c_{i,\sigma}$. 
${\cal H}$ consists of 
\begin{itemize}
\item a hopping term $-t$ that couples nearest-neighbor sites, and 
accounts for the quantum kinetic energy,
\item the pairwise electron-electron interaction, which itself 
consists of a $2U$ Hubbard repulsion when two electrons are at 
the same site $i$ with opposite spins and a $U/|i-i'|$ spin 
independent Coulomb repulsion when they are separated by a 
distance $|i-i'|$ (smallest distance between the sites $i$ 
and $i'$ on a square lattice with periodic BCs),
\item on site random potentials $v_i$ which are uniformly distributed 
inside the interval $[-W/2,W/2]$.
\end{itemize}
The {\it clean} system is obtained when the disorder strength $W$
is set to zero. In our model with a lattice spacing $a$, 
$\hbar^2/(2m^*a^2)\to t$, $e^2/(\epsilon a)\to U$, such that the 
factor $r_s$ becomes:
\begin{equation}
r_s=\frac{1}{\sqrt{\pi n_s} a^*_B} = \frac{U}{2t\sqrt{\pi n_e}}, 
\end{equation} 
for a filling factor $n_e=N/L^2$. This dimensionless ratio 
$r_s$ will allow us to compare our results obtained as a function 
of $U$ for a fixed filling factor $n_e$ and the experimental results 
obtained as a function of $n_s$. 

We denote ${\cal S}$ and ${\cal S}_z$ the total spin and its component 
along an arbitrary direction $z$. Since $[{\cal S}^2,{\cal H}]=
[{\cal S}_z,{\cal H}]=0$, ${\cal H}$ can be written in a block-diagonal 
form, with $N+1$ blocks where $S_z=-N/2,\ldots,N/2$ respectively. 
When $B=0$, there is no preferential direction and the groundstate energy 
$E_0$ does not depend on $S_z$. For a groundstate of total spin $S$, 
${\cal H}$ has $2S+1$ blocks with the same lowest eigenenergy $E_0 (S^2)$ 
since $ E_0(S^2)=E_0(S^2,S_z); \; \; \scriptstyle{S_z=-S,-S+1,...,S-1,S}$. 
Therefore, the number $N_b$ of blocks of different $S_z$ and of 
same lowest energy gives the total spin $S=(N_b-1)/2$ of the groundstate. 
 
\begin{figure}
\centerline{  
\epsfxsize=10cm
\epsfysize=6.6cm
\epsfbox{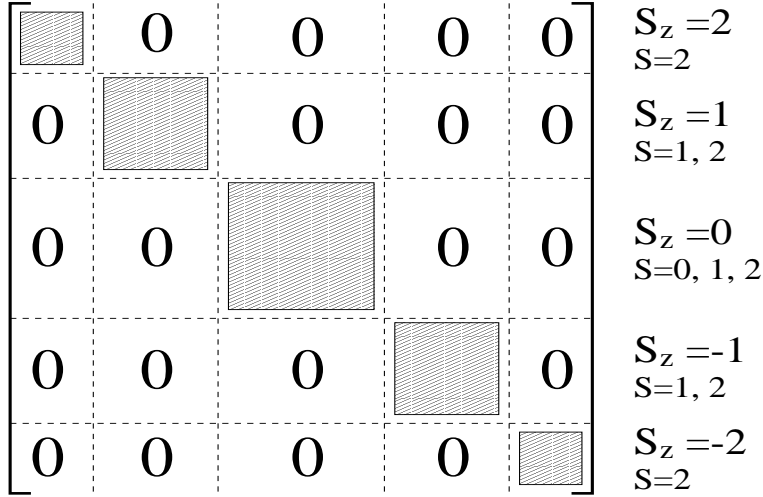}
}     
\caption{
Structure of the Hamiltonian matrix for 4 electrons in a $6 \times 6$ 
square lattice. The size of the different diagonal non zero sub-blocks 
are  $N(S_z)=396900, 257040$ and $58905$ for $S_z=0,\pm 1,\pm 2$ 
respectively.
}
\label{fig-m1}
\end{figure}

 If $N$ and $L$ are small enough, the ground state and the first 
excitations can be exactly calculated using Lanczos algorithm. 
Otherwise, certain approximations are unavoidable. Let us focus 
on the case $N=4$ and $L=6$ where the structure of ${\cal H}$ 
is given in Fig. \ref{fig-m1} for different $S_z$.  
Without magnetic field, we have the symmetry $\pm S_z$, and 
we have only to diagonalize the three sub-blocks with $S_z \geq 0$. 
${\cal H} (S_z=2)$ corresponds to fully polarized electrons (spinless 
fermions) where the orbital part of the wave-functions is totally 
anti-symmetric.

\section{Studied quantities}

 From exact diagonalization for small systems and using approximations 
for larger systems and weak coupling, we will study: 

\begin{itemize}

\item  the lowest eigenenergies $E_n(S_z)$ of the $S_z$ sub-blocks 
and the corresponding eigenvectors $|\Psi_n(S_z)>$. $n=0,1,2, \ldots$ 
corresponds to the states ordered by increasing energies. 

\item The local persistent currents $\vec{J}(i)$ created at a site $i$ 
by an Ahronov Bohm flux $\phi$ which is enclosed 
along the longitunal $l$-direction as sketched in Fig. \ref{fig-m2}. 
\begin{figure}[ht]  
\vskip.2in
\centerline{\epsfxsize=8cm\epsffile{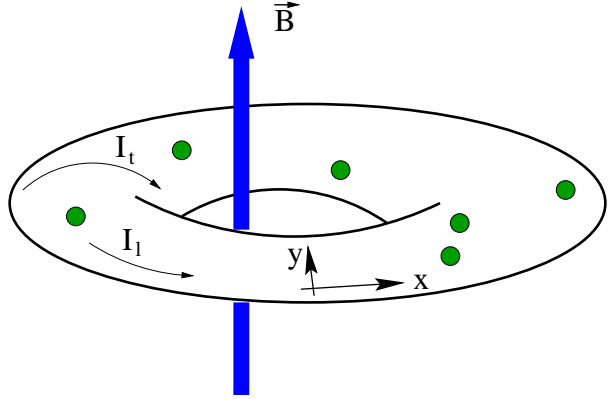}} 
\caption{2D Torus with $N$ electrons enclosing an Aharonov-Bohm flux 
$\phi = BL_x^2$} 
\label{fig-m2}
\end{figure} 
The flux $\phi$ can be included by taking appropriate longitudinal BCs 
(antiperiodic BCs corresponding to $\phi=\pi$ in our convention). 
The BCs along the transverse $t$-direction remain periodic.
The $\vec{J}(i)$ are vectors defined by their longitudinal and 
transverse components ($\vec{J}(i)=(J_{i,l},J_{i,t})$), or 
by their angles $\theta_i=\arctan(J_{i,t}/J_{i,l})$ 
and their absolute values $J_i=|\vec{J_i}|$. The longitudinal 
component $J_{i,l}$ of an eigenstate $|\Psi \rangle$ is defined as 
\begin{equation} 
J_{i,l}=2\hbox{Im}\langle\Psi|c_{i_x+1,i_y}^\dagger
c_{i_x,i_y}\exp(i\phi / L)|\Psi\rangle, 
\end{equation}
and $J_{i,t}$ is given by 
\begin{equation} 
J_{i,t}=2\hbox{Im}\langle\Psi|c_{i_x,i_y+1}^\dagger
c_{i_x,i_y}|\Psi\rangle.  
\end{equation}
The total current $I^{(n)}$ of the $n^{\rm th}$ many-body 
wavefunction $|\Psi_n\rangle$ of energy $E_n$ has a total longitudinal 
component $I_l(n)$ given by 
\begin{equation} 
I_l(n)(\overline{\phi})=-\left.\frac{\partial E_n}{\partial\phi} 
\right|_{\phi=\overline{\phi}}=\frac{\sum_{i}J_{i,l}(n)}{L}.
\end{equation} 
which will be calculated for $\overline{\phi}=\pi/2$. 

\item The crystallization parameter $\gamma$ defined using the 
function $C(r)=N^{-1} \sum_i \rho_i \rho_{i-r}$, where  
$\rho_i = \langle \Psi | n_i| \Psi \rangle$ is the electronic 
density of the state $|\Psi \rangle$ at the site $i$. 
The crystallization parameter $\gamma$ is given by 
\begin{equation} 
\gamma=\max_{\,r} C(r) - \min_{\,r} C(r)
\end{equation}
Note that $\gamma=1$ when the $N$ particles are localized on 
$N$ lattice sites and form a rigid  solid and $0$ when 
they are extended on the $L^2$ sites and form an 
homogenous liquid.

\item The participation ratio $\chi = N^2(\sum_i\rho_i^2)^{-1}$, 
 which gives the typical number of lattice sites occupied by an 
eigenstate $|\Psi>$.

\item The spectral parameter $\eta$ which characterizes the level 
repulsion. Uncorrelated spectra exhibit Poisson statistics. 
Correlated spectra can be described by Random Matrix Theory 
with Wigner-Dyson (W-D) statistics. For the one body spectra, 
the distribution $P(s)$ of the normalized 
energy spacings between consecutive levels has two different forms 
when $L \rightarrow \infty$:  the Poisson distribution 
$P_P(s)=\exp(-s)$ if the wavefunctions are localized, the Wigner 
surmises $P_W^O(s)=(\pi s/2)\exp(-\pi s^2/4)$ with time reversal 
symmetry (TRS) and $P_W^U(s)=(32 s^2/\pi^2)\exp(-4 s^2/\pi)$ without 
TRS, if the wave functions are extended. For the normalized $N$-body 
energy spacings 
$s_n=(E_{n+1}-E_n)/<E_{n+1}-E_n>$ (the brackets denote ensemble 
average), we define a spectral parameter:  
\begin{equation}
\eta_{(O,U)}=\frac{\hbox{var}(P(s))-\hbox{var}(P_W^{(O,U)}(s))}   
{\hbox{var}(P_P(s))-\hbox{var}(P_W^{(O,U)}(s))},
\end{equation}
($\eta_O$ with TRS, $\eta_U$ in the absence of TRS, for instance 
when $\phi =\pi/2)$). $\hbox{var}(P(s))$ denotes the variance of 
$P(s)$. The spectral parameter $\eta=1$ when $P(s)=P_P(s)$ and 
$\eta_{(O,U)}=0$ when $P(s)=P_W^{(O,U)}(s)$. 

\item The Zeeman energy necessary to polarize a non magnetized 
cluster. A {\it parallel} magnetic field $B$ does not induce 
orbital or Aharonov-Bohm effects, but defines the $z$-direction and removes 
the $S_z$ degeneracy by the Zeeman energy $-g\mu B S_z$. The 
ground state energy and its magnetization are given by the 
minimum of $E_0(S^2,S_z,B=0)-g\mu B S_z $. For a $S=0$ groundstate 
without field, the value $B^*$ for which $E_0(S_z)-g\mu B^* S_z 
\textstyle{=E_0(S_z=0)}$ defines the field  necessary to polarize 
the system to $S\geq S_z$. 
If one studies $N=4$ electrons, the total  
$Q_2=E_0(S_z=2)-E_0(S_z=0)$ and partial 
$Q_1 =E_0(S_z=1)-E_0(S_z=0)$ polarization energies give the 
Zeeman energies necessary to yield $S=2$ and $S=1$ respectively for 
a cluster with $S=0$. 

\end{itemize}

\section{Intermediate coupling regime for spinless fermions and weak disorder}

 We first consider an ensemble of disordered clusters with $L=6$ and $N=4$. 
The ground state (GS) and the first excitations 
of the fully polarized sub-block ($S_z=2$, spinless fermions) of the 
Hamiltonian matrix shown in Fig.\ref{fig-m1} have been obtained using the 
Lanczos algorithm. The statistical ensemble typically includes $10^3-10^4$ 
samples obtained from a disorder distribution with $W=5$. This is a relatively 
weak disorder for which one has quantum diffusion when $r_s=0$ inside the 
small clusters (no Anderson localization).

\subsection{Ground State}

\begin{figure}[ht]  
\vskip.2in
\centerline{\epsfxsize=9cm\epsffile{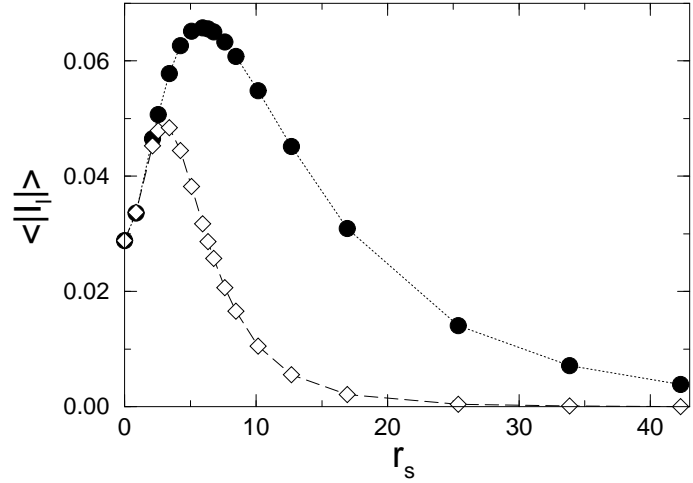}} 
\caption{Ensemble average longitudinal GS current $<I_l>$  
as a function of $r_s$ for $N=4$, $L=6$ and $W=5$. Exact 
values (filled symbols) and HF values (empty symbols). 
}
\label{fig-HF} 
\end{figure}

We summarize in this subsection the main results published in 
Ref. \cite{prl} and complementary unpublished results. As one 
switches on $U$, a first characteristic threshold $r_s^F$ 
can be identified by looking at the average total longitudinal 
persistent current $I_l$ of the GS at $\phi=\pi/2$, and comparing the 
exact quantity with the Hartree-Fock (HF) approximation (see appendix). 
Below $r_s^F \approx 5$, the mean field approximation reproduces 
the exact $I_l$, but strongly underestimates $I_l$ 
above $r_s^F$. This sharp breakdown of the HF approximation 
shown in Fig. \ref{fig-HF} 
means that strong correlation effects occur above $r_s^F$, such 
that the shift the GS energy when the BCs are changed cannot be 
obtained assuming the best possible SD for the ground state.  

\begin{figure}[ht]  
\vskip.2in
\centerline{\epsfxsize=4cm\epsffile{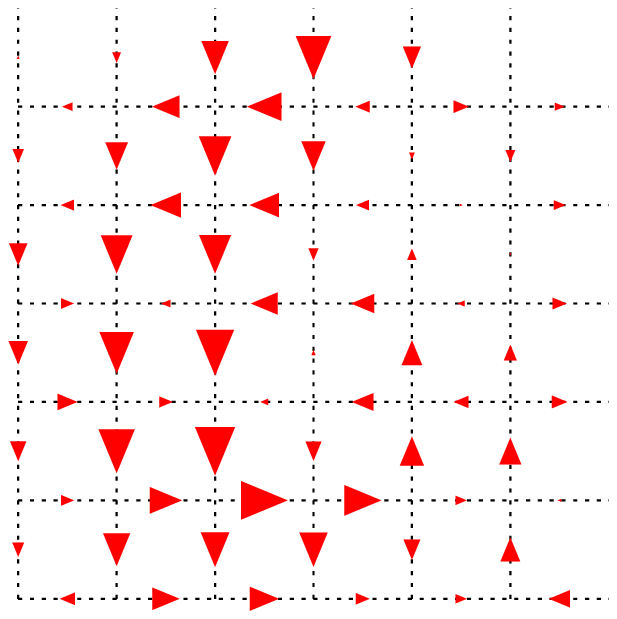} 
\epsfxsize=4cm\epsffile{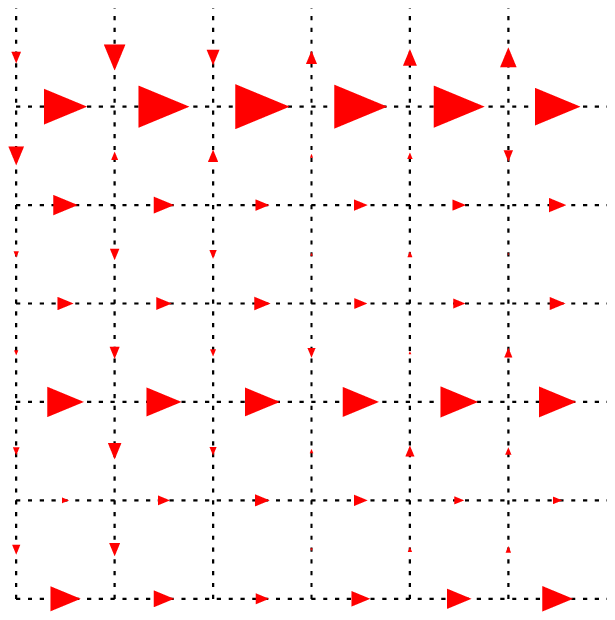} 
\hfill\epsfxsize=4cm\epsffile{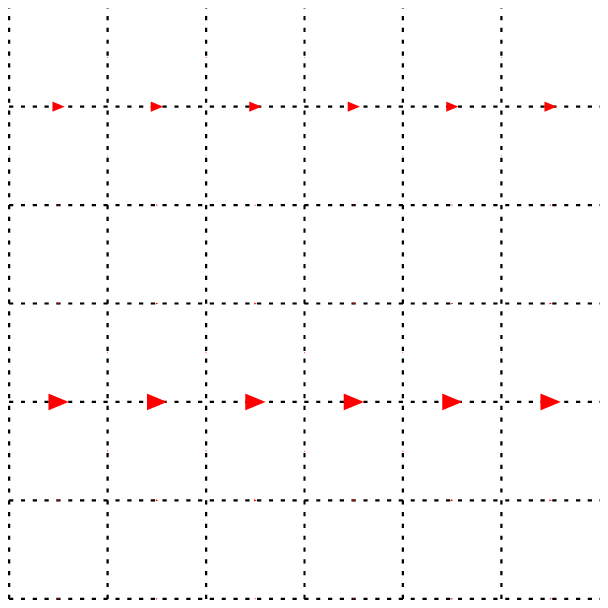}} 
\caption{Map of the local persistent currents in a given sample 
for small (left), intermediate (center) and large (right) 
values of $r_s$. 
}
\label{fig-carte} 
\end{figure}

A closer investigation of the persistent currents on a typical 
sample gives three regimes, as shown in Fig.\ref{fig-carte}. 
See also Ref. \cite{avishai-berkovits}.
For weak coupling, the local currents flow randomly inside the 
cluster, due to elastic scattering on the site potentials. 
For intermediate coupling, the pattern of the persistent currents 
becomes oriented along the shortest direction enclosing $\phi$. 
For large coupling, the oriented currents vanish. 
Ref. \cite{selva-weinmann} gives a detailed study of the large 
coupling limit where one can use perturbation theory for having 
the sign and the magnitude of $I_l$. 
 
\begin{figure}[ht]  
\vskip.2in
\centerline{\epsfxsize=12cm\epsffile{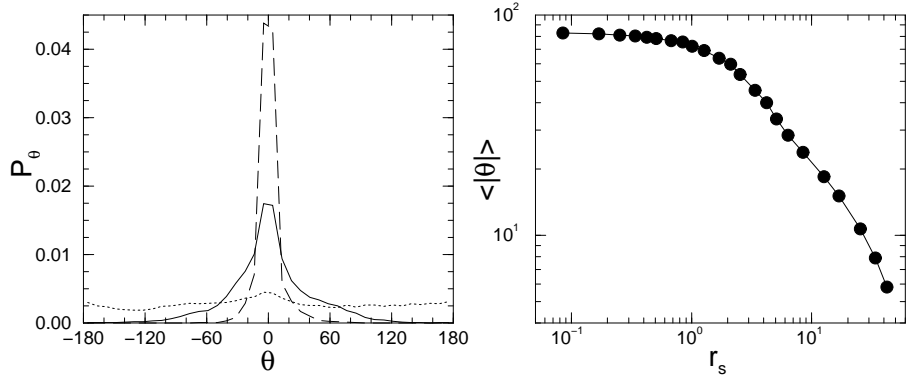}} 
\caption{Left: Distribution $P_{\theta}$ of the GS  
local current angles for $r_s=0$ (dotted line), 
$r_s=6.3$ (full line) and $r_s=42$ (dashed line). 
Right: Ensemble average angle $<|\theta|>$ as a 
function of $r_s$. 
} 
\label{fig-theta}
\end{figure}

If one looks at the distribution of the angles $\theta_i$ 
of the local currents, one can see in Fig.\ref{fig-theta} 
that the currents are randomly scattered without interaction, 
and that they become aligned when one goes to the strong 
coupling limit. The ensemble average value $<|\theta|>$ 
allows us to quantify the progressive change. If $p(\theta)=1/(2\pi)$, 
$<|\theta|>=\pi/2$, a value obtained for the low ratios $r_s$. 
At large $r_s$, $<|\theta|> \rightarrow 0$. The ratio $r_s$ at 
which the local currents cease to be oriented at random is 
consistent with the critical ratio $r_s^F$ where the HF approximation 
breaks down.
  
\begin{figure}[ht]  
\vskip.2in
\centerline{\epsfxsize=9cm\epsffile{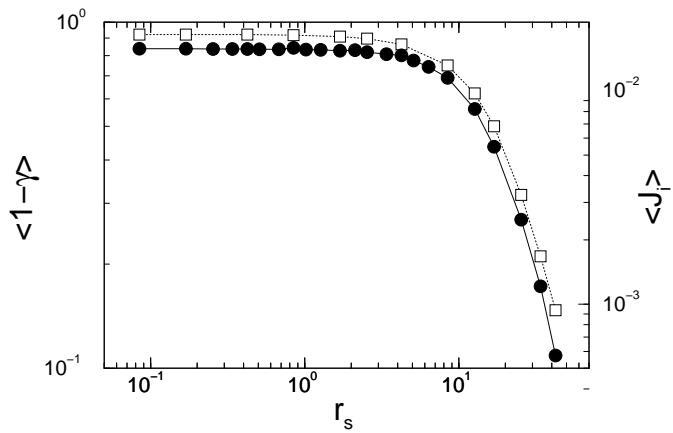}} 
\caption{Averages of the GS crystallization parameter $<1-\gamma>$ 
(left scale, empty symbols) and GS local current amplitude $<J_i>$ 
(right scale, filled symbols) as a function of $r_s$.} 
\label{fig-J-gamma} 
\end{figure}

By studying the average amplitude of the local
currents, one can see in Fig. \ref{fig-J-gamma} that  $<J_i>$ 
is essentially independent of $r_s$ up to a second threshold 
$r_s^W \approx 10$ which exceeds $r_s^F$. Moreover, comparing 
in Fig. \ref{fig-J-gamma} the GS average crystallization 
parameter $<\gamma>$ and $<J_i>$, on can see that the suppression 
of the persistent currents coincides 
with the formation of a solid Wigner molecule inside the disordered 
clusters. 

 The response of the ground state to an enclosed Aharonov-Bohm flux 
shows us that an intermediate correlated regime takes place between 
the Fermi limit and the Wigner limit, when typically $5<r_s<10$.  

\subsection{Low energy excitations}

In Ref. \cite{epl2}, the low energy excitations of the same
clusters have been studied, notably their statistics when the microscopic 
configurations of the random substrate are changed. For intermediate 
ratios $r_s$, the GS and the $8$ first following low energy excitations 
are characterized by oriented non random persistent currents and do not 
exhibit Wigner-Dyson (W-D) spectral statistics. Above those states, 
when the excitation energy $\epsilon$ exceeds an energy of the order of 
the Fermi energy $\epsilon_F$, the local currents become 
randomly oriented and the levels obey W-D statistics. Incidentally, let us 
note that the metallic behavior observed for intermediate couplings  
disappears also when the temperature exceeds a temperature of 
the order of the Fermi temperature. 

\begin{figure}[ht]  
\vskip.2in
\centerline{\epsfxsize=12cm\epsffile{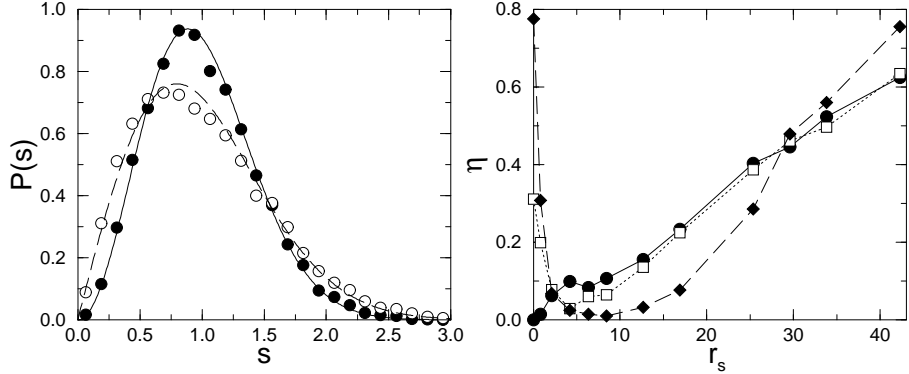}} 
\caption
{
Left: Spacing distribution $P(s)$ for $r_s=6.3$ 
when $\phi=0$ ($\circ$) and $\phi=\pi/2$ ($\bullet$) 
for energy levels above the 9 first levels 
(excitation energies $1.4  < \epsilon/\epsilon_F < 1.9$), 
compared to $P_O^W(s)$ (dashed line) and 
$P_U^W(s)$ (continuous line). Right: Spectral parameter $\eta_U$ 
as a function of $r_s$ for the first excitation $s_0$ (circles), 
$s_2-s_4$ (squares) and $s_{10}-s_{20}$ (diamonds).
}
\label{fig-statistics}
\end{figure} 

 In Fig. \ref{fig-statistics} (left) one can see that the spacing 
distribution $P(s)$ calculated for $r_s=6.3$ using the low energy 
levels except the 9 first levels ($n=10, \ldots ,20$), is given by 
the Wigner surmise, with an orthogonal-unitary crossover when one 
turns on an Aharonov-Bohm flux $\phi =0 \rightarrow \pi/2$. This 
corresponds to excitation energies $ 1.4 < \epsilon/\epsilon_F < 1.9$. 
Taking $\phi=\pi/2$, the variation of the spectral parameter $\eta_U$ 
as a function of $r_s$ is given for the successive level spacings in 
Fig. \ref{fig-statistics}. 
The first excitation is described by the Wigner surmise ($\eta_U=0$) 
without interaction but becomes more and more Poissonian when $r_s$ 
increases. The spacings characterizing the levels with $n=10, \ldots ,20$ 
have an opposite behavior. For $r_s = 0$, those excitations being the sum 
of more than one single-electron excitation are essentially uncorrelated, but 
become correlated for intermediate $r_s$ ($\eta_U \approx 0$) before 
being again uncorrelated at larger $r_s$. 

In summary, when one considers the low energy spectral statistics, a 
complementary signature of an intermediate regime is obtained, given by 
W-D statistics and randomly oriented local persistent currents outside 
the $9$ first states for which the absence of W-D statistics for 
intermediate $r_s$ is accompanied by a non random orientation of the 
persistent current angles $\theta$ (see Ref. \cite{epl2}). This behavior 
does not appear for weak and strong couplings, where the low energy spectral 
correlations decrease as the excitation energy increases.

\subsection{Intermediate liquid-solid regime in the clean limit}
 
 The fact that the 9 first states do not display quantum ergodicity 
for intermediate coupling and weak disorder suggests the existence of 
$9$ low energy collective excitations. A collective motion cannot be 
due to impurity scattering and should come from the corresponding 
clean limit. This limit has been investigated in Ref. \cite{katomeris}. 

When $W=0$, one has a single system which remains invariant 
under rotation of angle $\pi/2$ and under translations and 
reflections along the longitudinal $x$ and transverse $y$ 
directions. Invariance under translations implies that the 
momentum $K$ is a good quantum number 
which remains unchanged when $U$ varies. The symmetries imply that 
the states are fourfold degenerate if $K \neq 0$ and can be non 
degenerate if $K=0$. 

%
%

 When $U=0$, the states are $N_H$ plane wave Slater determinants (SDs)  
$d^{\dagger}_{k(4)}d^{\dagger}_{k(3)}d^{\dagger}_{k(2)}d^{\dagger}_{k(1)} 
|0>$, where $d^{\dagger}_{k(p)}$ creates a particle in a state of momentum 
$k(p)= 2\pi (p_x, p_y)/L$ ($p_{x,y}=1, \ldots, L$) and $|0>$ is the vacuum 
state. For $N=4$ and $L=6$, $N_H=58905$. The low energy eigenstates are 
given by the following plane wave SDs:

\begin{itemize}

\item $4$ degenerate ground states (GSs)  
$|K_0(\beta)>$ ($\beta=1,\ldots,4$) of 
energy $E_0(U=0)=-13 t$ and of momenta  $K_0=(0,\pm \pi/3)$ and 
$(\pm \pi/3,0)$.

\item  $25$ first excitations of energy $E_1 (U=0)=-12 t$ out of which 
$4$ plane wave SDs $|K_1(\beta)>$ will play a particular role for 
describing the intermediate GS. They 
correspond to a particle at an energy $-4t$ with $k(1)=(0,0)$, two 
particles at an energy $-3t$ and a fourth particle of energy $-2t$ 
with momenta such that $\sum_{j=2}^4 k(j)=0$. One has $k(2)=(0,\pm \pi/3)$, 
$k(3)=(\pm \pi/3,0)$ and $k(4)= (\mp \pi/3,\mp \pi/3)$ or $k(2)=(0,\mp 
\pi/3)$, $k(3)=(\pm \pi/3,0)$ and $k(4)=(\mp \pi/3,\pm \pi/3)$.

\item $64$ second excitations $|K_2(\alpha)>$ of energy $E_2(U=0)=-11t$.

\item $180$ third excitations $|K_3(\alpha)>$ of energy $E_3(U=0)=-10 t$. 

\item $384$ fourth excitations of energy $E_4(U=0)=-9t$ out of which 
$16$ plane waves SDs $|K_4(\delta)>$ will play a particular role 
for describing the intermediate GS. They are given by the condition 
that the total momentum is zero, which selects $20$ SDs out of which 
$4$ where the single particle state of energy $-3t$ is not occupied 
do not contribute. The $|K_4(\delta)>$ are $16$ SDs of energy $-9t$, 
given by $8$ SDs where the particles have energies $-4t,-3t,-2t,0t$ 
respectively and by $8$ other SDs where the particles have energies 
$-3t,-3t,-2t,-t$ respectively.

\end{itemize} 

When $t=0$, the states are $N_H$ Slater determinants 
$c^{\dagger}_ic^{\dagger}_jc^{\dagger}_kc^{\dagger}_l |0>$ built out 
from the site orbitals. The configurations $ijkl$ correspond to the $N_H$ 
different patterns characterizing $4$ different sites of the $6 \times 6$ 
square lattice.  The low energy part of the spectrum is made of the 
following site SDs:

\begin{itemize}

\item $9$ squares $|S_0(I)>$ ($I=1, \ldots, 9$) of side 
$a=3$ and of energy $E_0(t=0) \approx 1.80 U$.
 
\item $36$ parallelograms $|S_1(I)>$ of sides ($3,\sqrt{10}$) and of energy 
$\approx 1.85 U$.

\item $36$ other parallelograms $|S_2(I)>$ of sides ($\sqrt{10}, \sqrt{10})$ 
and of energy $\approx 1.97 U$. 

\item $144$ deformed squares $|S_3(I)>$ obtained by moving a single site of a 
square $|S_I>$ by one lattice spacing and of energy $\approx 2 U$. 

\end{itemize} 
 
%
%

\begin{figure}[ht]  
\vskip.2in
\centerline{\epsfxsize=12cm\epsffile{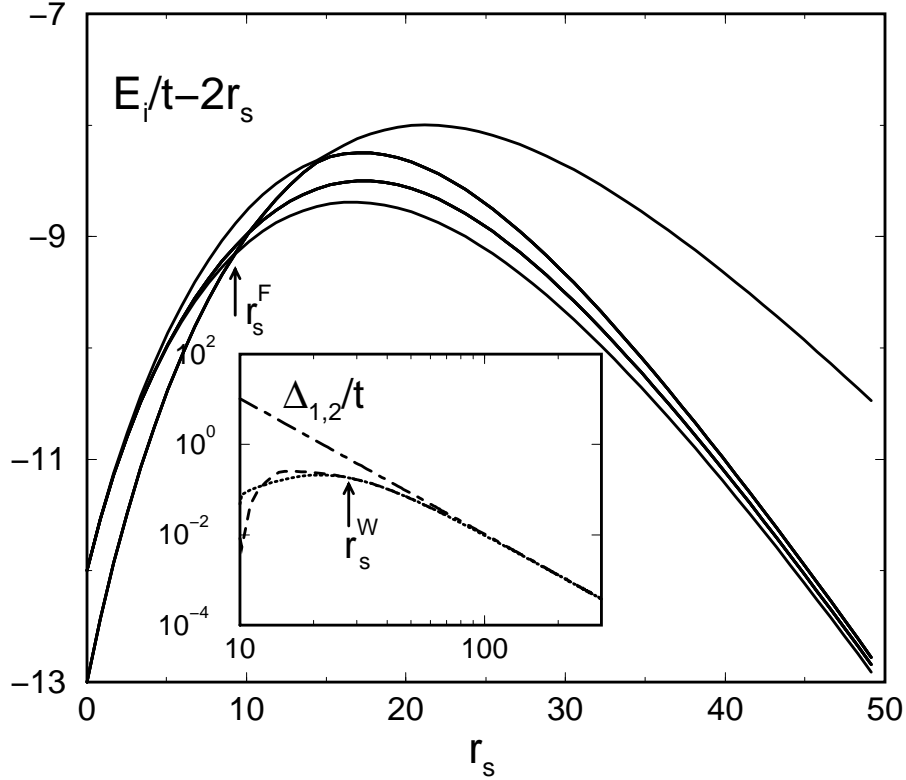}} 
\caption{
As a function of $r_s$, low energy part of the spectrum 
exhibiting a GS level crossing at $r_s^F$. Inset: two first 
level spacings $\Delta_1/t$ (dashed) and $\Delta_2/t$ (dotted) 
which become equal at $r_s^W$ and the perturbative result 
$\Delta_1/t=\Delta_2/t \approx 10392/ r_s^{3}$ valid when 
$r_s \rightarrow \infty$ (dot-dashed).}
\label{W-AL-FIG1}
\end{figure}  

For the first low energy states, the crossover from the $U=0$ eigenbasis 
towards the $t=0$ eigenbasis is shown in Fig. \ref{W-AL-FIG1} when one 
increases the ratio $r_s$. If we follow the $4$ GSs $E_0(r_s=0)$ 
($K_0\neq 0$), one can see a first level crossing at $r_s^F \approx 9.3$ 
with a non degenerate state ($K_0=0$) which becomes the GS above $r_s^F$, 
followed by two other crossings with two other sets of $4$ states with 
$K_I \neq 0$. When $r_s$ is large, $9$ states coming from $E_1(r_s=0)$ 
have a smaller energy than the $4$ states coming from $E_0(r_s=0)$. The 
degeneracies ordered by increasing energy become $(1,4,4,4,\ldots)$ 
instead of $(4,25,64,\ldots)$ for $r_s=0$. Since the degeneracies are 
$(9,36,36,\ldots)$ when $t=0$, these $9$ states give the $9$ 
square molecules $|S_0(I)>$ when $r_s \rightarrow \infty$. When 
$r_s^{-1}$ is very small, the first $9$ states correspond to a single 
massive molecule free to move on a restricted $3 \times 3$ lattice,
the single non frozen degree of freedom in this limit being the location 
$R_I$ of the center of mass of the $|S_0(I)>$. One has an effective hopping 
term $T\propto t r_s^{-3}$ when $N=4$ and the total momentum is quantized 
($K_l(I)=2\pi p_l/3$ and  $K_t(I)=2\pi p_t/3$ being its longitudinal and 
transverse components respectively with $p_{l,t}=1,2,3$). For a square 
lattice at a filling factor $1/9$, the $R_I$ are indeed located on a periodic 
$3 \times 3$ square lattice. This is an important simplification of 
our model. This gives $9$ states of  
kinetic energies given by $-2T (\cos K_l(I) +\cos K_t(I))$. The kinetic part 
of the low energy spectrum is then $-4T,-T,+2T$ with degeneracies $1,4,4$ 
respectively. This structure with two equal energy spacings 
$\Delta_1$ and $\Delta_2$ appears (inset of Fig. \ref{W-AL-FIG1}) when $r_s$ 
is larger than the crystallization threshold $r_s^W \approx 28$. Above 
$r_s^W$, to create a defect in the rigid molecule  costs a high energy 
available in the $10 ^{th}$ excitation only. We have seen in the 
previous section that the $9$ first levels do not obey Wigner-Dyson 
statistics at intermediate $r_s$ when a random potential 
is added, in contrast to the following levels. The study of the 
clean limit gives us the explanation. The two characteristic thresholds 
$r_s^F$ (level crossing) and $r_s^W$ (9 first states having the 
structure of the spectrum of a single massive molecule free to move 
on a $3 \times 3$ square lattice) can also be detected by other methods 
given in Ref. \cite{katomeris}.  

\begin{figure}[ht]  
\vskip.2in
\centerline{\epsfxsize=10cm\epsffile{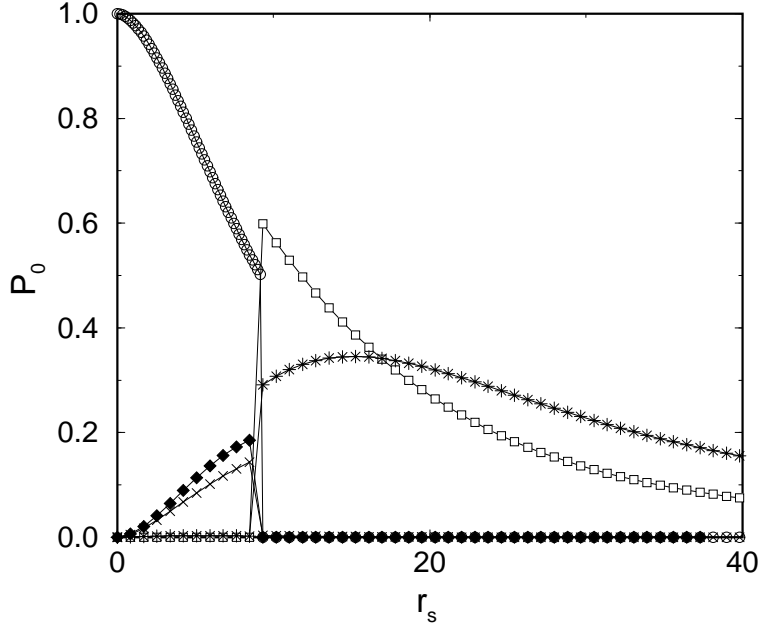}} 
\caption{Ground state projections $P_0(r_s)$ onto a few plane wave SDs, 
given by the $4 |K_0(\beta)>$ (empty circle), the 4 $|K_1(\beta)>$ 
(empty square), the 64 $|K_1(\alpha)>$ (filled diamond), the 180 
$|K_2(\alpha)>$ ($\times$), the 16 $|K_4(\delta)>$ (asterisk) 
respectively, as a function of $r_s$. } 
\label{FIG-P-zero}
\end{figure}

To understand further the nature of the intermediate GS, we have 
projected the GS wave functions $|\Psi_0(r_s)>$ over the 
low energy eigenvectors of the two eigenbases valid for 
$U/t=0$ (Fig. \ref{FIG-P-zero}) and for 
$t/U=0$ (Fig. \ref{FIG-P-infty}) respectively. 

Let us begin by studying the GS structure in the $U=0$  
eigenbasis. Below $r_s^F$, each of the 4 GSs $|\Psi_0^{\alpha}(r_s)>$ with 
$K_0\neq 0$ has still a large projection  
\begin{equation}
P_0(r_s,0)=\sum_{\beta=1}^4 |<\Psi_0^{\alpha}(r_s)| K_0({\beta})>|^2 
\end{equation}
over the $4$ non interacting GSs. There is no projection over the $25$ first 
excitations and smaller projections $P_0(r_s,2)$ and $P_0(r_s,3)$ over the 
$64$ second and $180$ third excitations of the non interacting system. 
Above $r_s^F$, the non degenerate GS with $K_0=0$ has a large projection 
\begin{equation}
P_0(r_s,1)=\sum_{\beta=1}^4 |<\Psi_0(r_s)| K_1({\beta})>|^2 
\end{equation}
which is equally distributed over the $4$ excitations $|K_1(\beta)>$ 
of momentum $K_1=0$ and a second significant contribution 
\begin{equation}
P_0(r_s,4)=\sum_{\delta=1}^{16} |<\Psi_0(r_s)| K_4({\delta})>|^2 
\end{equation}
given by its projection onto the $16$ plane wave SDs $|K_4({\delta})>$ 
belonging to the fourth excitation of the non interacting system. 
Above $r_s^F$, its projections onto the $4$ $|K_0({\beta})>$, the 
$21$ other first excitations and the second and third excitations of 
the non interacting system are zero or extremely negligible. 

\begin{figure}[ht]  
\vskip.2in
\centerline{\epsfxsize=10cm\epsffile{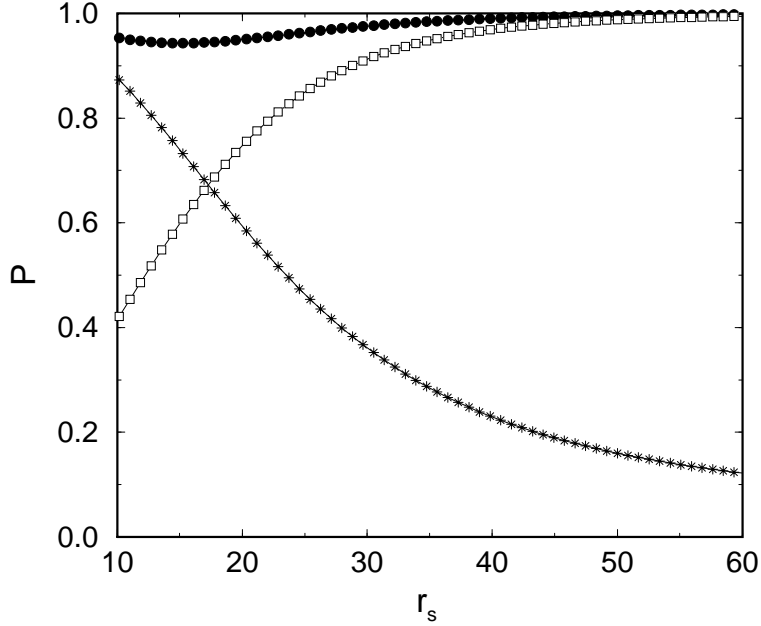}} 
\caption{Ground state projection $P_0^t(r_s)$ (asterisk) and 
$P_{\infty}^t(r_s)$ (empty square) onto the subspace spanned by 
the low energy plane wave and site SDs respectively,
and total GS projection $P$ (filled circle) onto the 
re-orthonormalized basis using the low energy eigenvectors 
of the two limiting bases.} 
\label{FIG-combined-bases}
\end{figure}

The total GS projection 
\begin{equation}
P_0^t(r_s)=P_0(r_s,1)+P_0(r_s,4) 
\end{equation}
onto the $4$ $|K_1({\beta})>$ and $16$ $|K_4({\delta})>$ is given 
in Fig. \ref{FIG-combined-bases} when $r_s > r_s^F$. This shows us 
that a large part of the system remains an excited liquid 
above $r_s^F$, given by a special rule of occupation of the one 
particle plane wave states. The occupation of the one body states is 
very different from the usual Pauli rule after the level crossing $r_s^F$. 
A necessary, but non sufficient condition for a plane wave SD to 
significantly contribute to the zero momentum GS is of course to have 
a zero total momentum. Those projections decrease as $r_s$ increases and 
become negligible in the strong coupling limit. A complete GS description 
in this limit will require more and more plane wave SDs.

\begin{figure}[ht]  
\vskip.2in
\centerline{\epsfxsize=10cm\epsffile{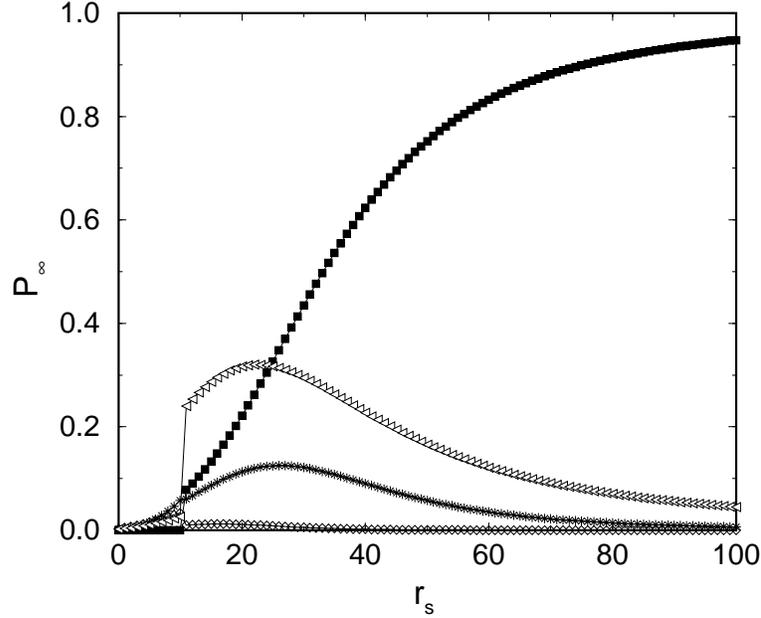}} 
\caption{Ground state projection $P_{\infty}(r_s)$ onto a few site SDs, 
given by the $9$ squares $|S_0(I)>$ (filled square), the $36$ parallelograms 
$|S_1(I)>$ (asterisk), the $36$ parallelograms $|S_2(I)>$ (diamond), 
and the $144$ deformed squares $|S_3(I)>$ (left triangle) respectively, 
as a function of $r_s$. } 
\label{FIG-P-infty}
\end{figure}

 We now study the GS projections $P_{\infty}$ onto the $t=0$ eigenbasis.  
The GS projection 
\begin{equation}
P_{\infty}(r_s,0)=\sum_{I=1}^9 |<\Psi_0^{\alpha}(r_s)| S_0(I)>|^2 
\end{equation}
onto the $9$ square site SDs $|S_0(I)>$ is given in Fig. \ref{FIG-P-infty}, 
together with the GS projection $P_{\infty}(r_s,J)$ onto the 
site SDs corresponding to the $J^{th}$ degenerate low energy excitations 
of the $t=0$ system.  The total GS projection 
\begin{equation}
P_{\infty}^t(r_s)= \sum_{p=0}^3 P_{\infty}(r_s,p) 
\end{equation}
onto the $9$ squares $|S_0(I)>$, the $36$ parallelograms $|S_1(I)>$, 
the $36$ other parallelograms $|S_2(I)>$ and the $144$ deformed squares 
$|S_3(I)>$ is given in Fig. \ref{FIG-combined-bases} when $r_s > r_s^F$. 
This shows us that the ground state begins to be a floppy solid also 
above $r_s^F$. When $r_s$ increases, $P_{\infty}(r_s,0)$ 
goes to one, and the ground state is a simple rigid square Wigner 
molecule.

The site SDs and plane wave SDs are not orthonormal. After 
re-orthonormalization, the total projection $P$ of $|\Psi_0(r_s)>$ 
over the subspace spanned by the $4$ $|K_1(\beta)>$ and $16$ $|K_4(\delta)>$ 
and $225$ site SDs of lower electrostatic energies ($9$ squares, $36+36$ 
parallelograms, $144$ deformed squares) are given in 
Fig. \ref{FIG-combined-bases}. One can see that $|\Psi_0(r_s)>$ is almost 
entirely located inside this very small part of a huge Hilbert space 
for intermediate $r_s$, spanned by no more than $245$ low energy SDs of 
different nature, and adapted to describe a solid entangled with an  
excited liquid.

 From the study of the GS projections emerges the conclusion that a 
minimal description of the intermediate GS requires to combine the 
low energy states of the two limiting eigenbases. In this sense, the 
GS is neither solid, nor liquid, but rather the quantum superposition 
of those two states of matter. This is strongly reminiscent of the 
conjecture proposed in Ref. \cite{AL} for the quantum melting of 
solid Helium in three dimensions. It suggests possible improvements 
of the trial GS to use for intermediate $r_s$ in variational or 
path integral quantum Monte Carlo approaches. Instead of using Jastrow 
wave functions improving the plane wave SDs for the liquid or the site 
SDs for the solid, or their nodal surfaces, it will be interesting 
to study if a combination of the two, with a more complex nodal structure, 
 and describing a solid-liquid regime does not lower the GS energy for 
intermediate $r_s$. A positive answer would confirm that an unnoticed 
intermediate solid-liquid phase does exist in the 
thermodynamic limit for fermionic systems in two dimensions. 

\subsection{Magnetic signature of the intermediate regime for 
weak disorder}

We return to the study of weakly disordered samples when 
the spin degrees of freedom are included. Their role 
and the consequences of an applied parallel magnetic field 
which align only the spins without inducing orbital effects, 
have been the subject of Ref.\cite{selva}. 
A statistical ensemble of matrices having the structure given in 
Fig. \ref{fig-m1} have been studied for $W=5$, $N=4$ and $L=6$ 
providing complementary signatures of a particular intermediate behavior. 

\begin{figure}
\centerline{
\epsfxsize=12cm
\epsffile{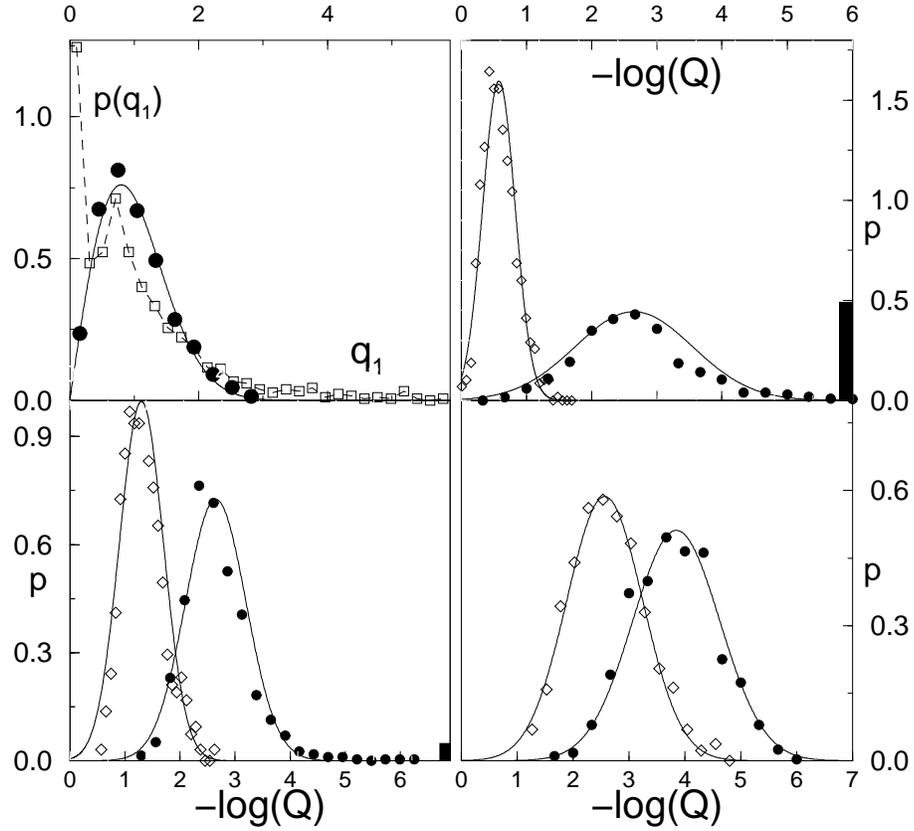}
}
\caption
{
Distributions $p$ of the polarization energies $Q_1$ and $Q_2$ 
at different values of $r_s$. Upper left: $p(q_1)$ at 
$r_s= 0$ (circle) and $2.5$ (square) where $q_1=Q_1/<Q_1>$. 
The continuous line is the Wigner surmise.  
$p(-\log Q_1)$ (filled circle) and $p(-\log Q_2)$ 
(empty diamond) 
at $r_s=2.5$ (upper right, right scale) $ 5.8 $ (lower left, left scale) and 
$16.8$ (lower right, right scale) respectively. The thick bars (put at right 
edge of the figures) give the peaks $\delta(Q_1)$ of the bimodal 
$p(Q_1)$. The continuous lines are normal fits.
}  
\label{magnetization-fig1bis}
\end{figure}

When $r_s=0$, the two one body states of lowest energy are doubly occupied 
and $S=0$ ($S=1/2$ if $N$ is odd). To polarize the $S=0$ ground state to 
$S=1$ corresponds to the transition of one electron at the Fermi energy 
and costs an energy equal to the one body level spacing. $p(Q_1)$ 
is then given by the spacing distribution $p(s)$ between consecutive 
one body levels, the Wigner surmise $P_W^O(s)$ in 
the diffusive regime. When $r_s$ is large, the $4$ electrons occupy 
the four sites $c_j$ $j=1,\ldots,4$ of the square configuration $|S_0(I)>$ 
of side $a=3$  with the lowest substrate energy $\sum_{j=1}^4 v_{c_j}$. 
The ground state in this limit becomes 
$|\Psi_c>=\prod_{j=1}^4 c^{\dagger}_{c_j,\sigma_j} |0>$ 
with a spin independent energy $E_c$. This square can support $2^N=16$ 
spin configurations. We summarize the main results of a perturbative 
expansion of $E_c$ in powers of $t/U$. The spin degeneracy of 
$E_c$ is removed by terms of order $t(t/U)^{2a-1}$, which is the 
smallest order where the $16$ spin configurations can be coupled via 
intermediate configurations allowing a double occupancy of the same 
site. Therefore, $2a-1$ is the order where the perturbation begins 
to depend on $S_z$ and $Q_1$ as $Q_2 \propto t(t/U)^{2a-1} 
\rightarrow 0$ when $t/U \rightarrow 0$ (we have numerically checked 
this decay when $r_s > 100$). Moreover, the correction  to $E_c$ 
depending on $S_z$ and $\propto t(t/U)^{2a-1}$  is given by an effective 
antiferromagnetic Heisenberg Hamiltonian. The $S=0$ ground state for large 
$r_s$ correspond to $4$ electrons forming an antiferromagnetic square Wigner 
molecule. However $Q_1$ and $Q_2$ are very small 
when $r_s$ is large, and the antiferromagnetic behavior can be an artefact 
due to the square lattice. Without impurities and in a continuous limit, 
a quasi-classical WKB expansion \cite{roger} valid for very large values 
of $r_s$ shows that 3 particle exchanges dominate, leading to 
ferromagnetism. Recent Monte-Carlo calculations 
\cite{bernu} suggest that the crystal in the continuous limit 
becomes a frustrated antiferromagnet closer to the melting point.

The perturbative corrections $\propto t(t/U)^{2a-1}$ depend 
on the random variables $v_i$ via $\prod_{J=1}^{2a-1} (E_c-E_J)^{-1}$ 
where the $E_J$ are the classical energies of the intermediate configurations. 
$E_J$ is the sum of an electrostatic energy and of a random substrate energy 
$E_s(J)=\sum_{k=1}^4 v_{J(k)}$. Due to the high order $2a-1$ of the 
correction, a normal distribution for $E_J$ leads to a log-normal 
distribution for $\prod_{J=1}^{2a-1} (E_c-E_J)^{-1}$. Therefore 
$p(\Delta_1)$ and $p(\Delta_2)$ should be log-normal when $r_s$ is large.  

 $p(Q_1)$ is given in Fig. \ref{magnetization-fig1bis} for different 
$r_s$.  The expected Wigner surmise takes place for $r_s=0$. 
A small interaction quickly drives $p(Q_1)$ towards a bimodal 
distribution, with a delta peak at $Q_1=0$ 
and a main peak centered around a non zero value of 
$Q_1$. The delta peak gives the probability to have spontaneously 
magnetized clusters with $S=1$. The main peak gives the field $B$ 
necessary to create $S=1$ in a cluster with $S=0$.  
The logarithmic scale used in Fig. \ref{magnetization-fig1bis} (upper right) 
underlines  the bimodal character of the distribution and confirms that 
the main peak becomes log-normal when $r_s$ is large. The distribution of 
$Q_2$ is not bimodal: a fully polarized cluster has never been seen 
when $B=0$. $Q_2$ becomes also log-normally distributed when $r_s$ 
is large.  

\begin{figure}
\centerline
{
\epsfxsize=10cm
\epsffile{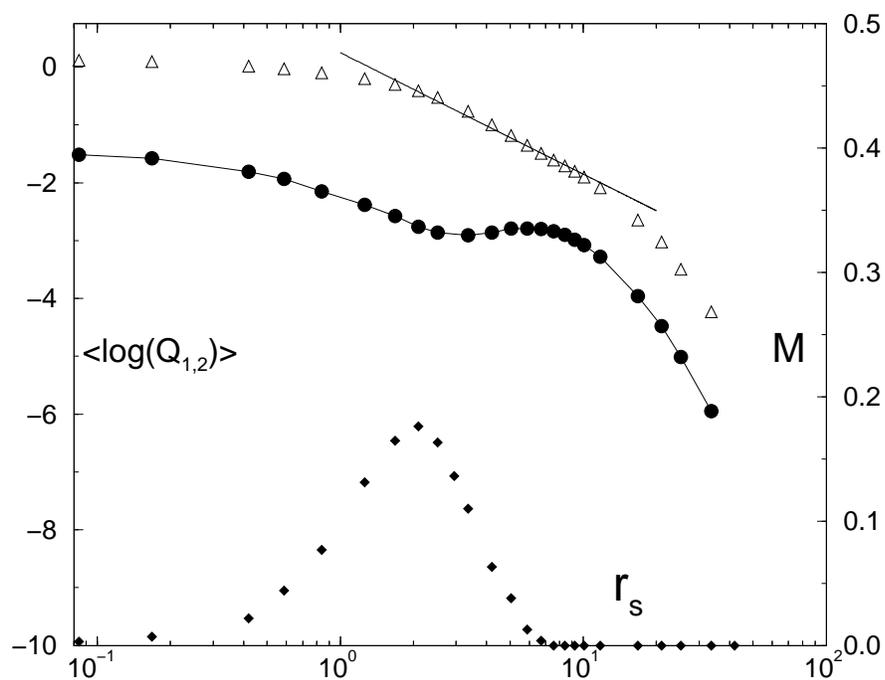}
}
\vspace{0.5cm}
\caption
{
As a function of $r_s$, fraction $M$ of clusters with $S=1$ 
at $B=0$ (filled diamond, right scale),  partial $<\log Q_1>$ 
(filled circle, left scale) and total $<\log Q_2>$ (empty triangle, 
left scale) energies required to polarize $S=0$ clusters to $S=1$ and $S=2$ 
respectively. The straight line corresponds to $0.25 - 2 \log r_s$. 
}
\label{magnetization-fig2}
\end{figure}

\begin{figure}
\centerline{
\epsfxsize=10cm
\epsffile{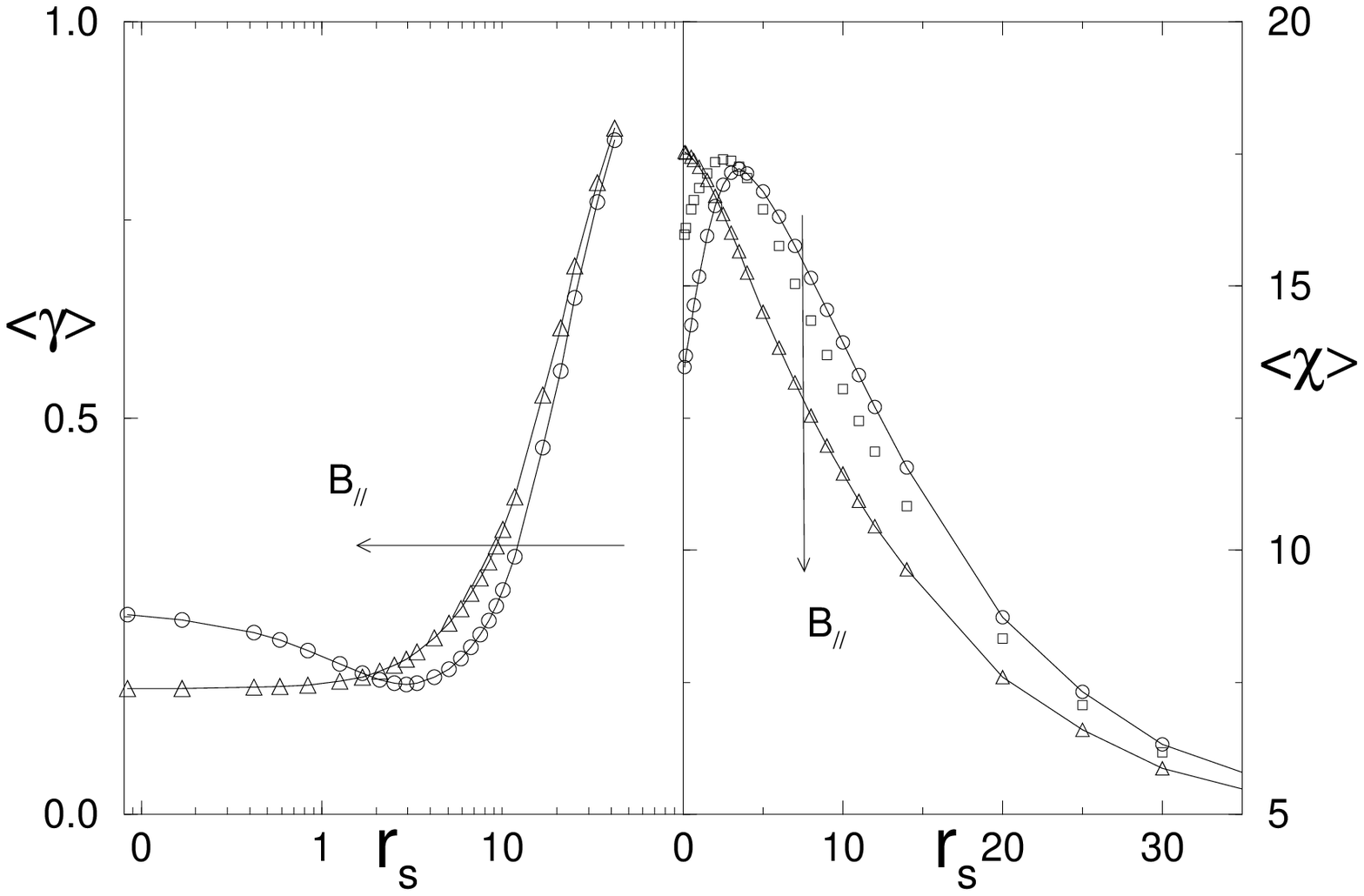}
}
\caption
{
As a function of $r_s$, ensemble averages of the crystallization parameter 
$<\gamma (S_z)>$ (right) and of the numbers of occupied sites $<\chi (S_z)>$ 
(left). $S_z=0$  (circle), $S_z=1$  (square) and $S_z=2$  (triangle).
The arrows indicate the effect of a parallel magnetic field.
}
\label{magnetization-fig3}
\end{figure}

 In Fig. \ref{magnetization-fig2}, the fraction $M$ of clusters with 
$S=1$ at $B=0$ is given as a function of $r_s$. One can see the mesoscopic 
Stoner instability (see appendix) taking place at $r_s \approx 0.35$. 
The Stoner mechanism should eventually give fully polarized electrons. 
This is not the case, the increase of $M$ breaks down when 
$r_s= r_s^{FS} \approx 2.2$, a value where the Stoner mechanism and 
hence the HF approximation break down. In the same clusters, we have seen 
that the HF approximation fails to describe the persistent currents of 
the fully polarized sub-block of ${\cal H}$ (spinless fermions) 
when $r_s> r_s^{FP} \approx 5$. $r_s^{FP}$ takes a smaller value 
$r_s^{FS}$ when the spin 
degrees of freedom are included. Above $r_s^{FS}$, 
$M$ regularly decreases to reach a zero value for $r_s^{WS} \approx 9$ 
where an antiferromagnetic square molecule is formed. In the intermediate 
regime, there is a competition between the Stoner ferromagnetism and the 
Wigner antiferromagnetism. Since the $S=0$ clusters are characterized 
by log-normal distributions, the ensemble averages $<\log Q_1>$ and 
$<\log Q_2>$ (without taking into account the $S=1$ spontaneously 
magnetized clusters) define the typical fields $B$ necessary to 
yield $S=1$ or $S=2$ in a $S=0$ cluster. Fig. \ref{magnetization-fig2} 
provides two magnetic signatures confirming the existence of a novel 
intermediate regime between the Fermi glass $(r_s < r_s^{FS})$ and the 
Wigner glass $(r_s>r_s^{WS})$: $<\log Q_1>$ becomes roughly independent 
of $r_s$, while $Q_2 \propto r_s^{-2}$. 

The number of occupied sites $\chi(S_z)$ depends on $S_z$ for small $r_s$ 
and becomes independent of $S_z$ for large $r_s$. At $r_s=0$, 
$\Psi_0(S_z=2,1,0)$ occupy respectively $4,3,2$ one body states while 
the Wigner molecule occupies $4$ sites only at large $r_s$. The ensemble 
average $<\chi(S_z=2)>$ shown in Fig. \ref{magnetization-fig3} is maximum 
when $r_s=0$ and decays as $r_s$ increases, suggesting the absence of 
delocalization for the polarized system. The non polarized system 
behaves differently, since $<\chi(S_z=1,0)>$ first increase to reach a 
maximum $\approx <\chi(S_z=2,r_s=0)>$ before decreasing. In 
Fig.\ref{magnetization-fig3} one can see also from the curves $\gamma(r_s)$ 
that charge crystallization is easier when the clusters are polarized than 
otherwise. The arrows indicated in Fig. \ref{magnetization-fig3} 
underline two consequences of a parallel field $B$: smaller number of 
occupied sites and smaller crystallization threshold.

\subsection{Numerics versus experiments}

 One can question whether the comparison between numerical simulations based 
on a small lattice model with a few particles and measures using $10^{11}$ 
electrons per $cm^2$ makes sense. It is nevertheless interesting to point 
out certains analogies in the obtained thresholds and behaviors. 

\begin{itemize}

\item The ratios $r_s^W$ where the rigid Wigner molecule is formed 
in small clusters are of the order of the ratios $r_s^c$ where 
one observes the MIT. One finds a value $\approx 10$ for a weakly 
disordered cluster which increases to larger values when one goes 
to the clean limit ($r_s^W\approx 30-40$). The $r_s^W$ numerically 
obtained for larger disorder ($W=5, 10, 15$) in Ref.\cite{prl} are 
roughly independent of $W$ when $W>5$ and reproduce the flat part of the 
experimental curve sketched in Fig. \ref{fig4}.

\item The large magnetoresistance sketched in Fig. \ref{fig5} is 
consistent with the reduction of $\chi$ yielded by a parallel magnetic 
field as indicated in Fig. \ref{magnetization-fig3} (right).

\item  The shift of the critical threshold $r_s^W$ when one polarizes 
the electrons with a parallel magnetic field (Fig. \ref{magnetization-fig3} -
left) is consistent with the shift of the critical densities sketched in 
Fig. \ref{fig3}.

\item A polarization energy $Q_2 \propto r_s^{-2}$ is consistent 
with the $n_s$ dependence of $B_{sat}$ given in Fig. \ref{fig5} 
for similar values $3 < r_s < 9$ \cite{shashkin1}. 

\item A threshold at $r_s^F \approx 3-5$ given by the study of the persistent 
currents and of the magnetization corresponds to the density at which 
the metallic behavior ceases to occur, according to Ref. \cite{hamilton}. 

\item It can be argued that the usual small negative magnetoresistance 
yielded by a perpendicular magnetic field for intermediate $r_s$ 
(Fig. \ref{fig6}) would mean that transport is due to the gapless 
excitations of the intermediate floppy quantum solid. The propagation of 
such excitations could be reduced by the usual weak localization 
corrections when one has elastic scattering by impurities.   

\end{itemize}

\subsection{Mesoscopic 2D electron systems confined in harmonic traps}

 For the important issue of the melting of a macroscopic 2D Wigner 
glass (crystal without disorder) and a possibly associated 2D-MIT, 
the study of finite size systems can give hints, but requires to be 
complemented by a systematic study of the scale dependence of the 
finite size effects at a given electron density.
Before reviewing some results where such a finite size scaling analysis 
has been done, let us underline that the formation of the mesoscopic Wigner 
molecule in a few electron system is by itself an important issue. 
One can use a few electrons \cite{dot} confined in a quantum dot or a few 
ions \cite{ion} trapped by electric and magnetic fields. Increasing 
the size of the trap, a crossover from independent-particle towards 
collective motion can be observed. Moreover, a 
localization-delocalization transition has been observed \cite{Ashoori} 
in a quantum dot using single-electron capacitance spectroscopy and increasing 
the number of trapped electrons. Considering electrons 
confined in a harmonic trap, and not on a 2D torus, other numerical 
studies have reached the same crucial conclusion, namely that between 
the Fermi system and the ``classical'' rigid molecule, there is an 
intermediate regime of a floppy weakly formed Wigner molecule. This was 
shown for instance in an exact study of a two-electron artificial 
atom \cite{yannouleas} and in a Monte Carlo study \cite{filinov} of 
a few electron system. In Ref. \cite{filinov}, this new regime corresponds 
to particles having a radial ordering on shells without correlated intershell 
rotation. This was attributed to the special symmetry of the  
harmonic trap and to the imposed density gradient. Our studies show 
that mesoscopic Wigner crystallization takes also place in two stages 
when the particles are confined on a 2D torus, with a uniform electron 
density.

\section{Quantum phase transition for weak coupling and strong disorder}

 In this Section we consider the case of spinless fermions for 
strong disorder and low filling factors $N/L^2$. 
In contrast to the previously studied systems, 
the relevant length scale without interaction, the one-body localization 
length $L_1$, is not only smaller than the system size, but also smaller 
than the average distance between particles, $d\propto n_s^{-1/2}$. 
Truncated bases built out from single-particle wavefunctions or 
Hartree-Fock orbitals will be used to study larger systems. These 
approximations remain reliable when the interaction strenght $U$ 
does not exceed the hopping term $t$, but become uncontrolled in the 
strongly correlated regime $U\gg t$. They allow us to study 
the finite size effects in the limit of relatively weak coupling. 
They show the existence of a second order phase transition between 
the weak coupling Fermi limit and a new phase appearing above 
$U_c \approx t$ for $W=10-15$. Two numerical evidences of a transition 
are presented. 
The first is given by the divergence of a characteristic length 
of the two dimensional system which allows us to map the finite 
size data onto a single scaling curve. The second is given by the 
existence of a size independent distribution of the first excitation 
at $U_c$. 

In this limit, the use of the parameter $r_s$ becomes 
questionable, since it is no longer an appropriate measure 
of the Coulomb to Fermi energy ratio, the one body 
part of the Hamiltonian being deeply changed 
by the random substrate. The values $U_c$ obtained 
when $W$ is large give for the considered filling factors critical 
values $r_s^c = U_c/(2t \sqrt{\pi (N/L^2)}$ of the order 
of the previously obtained ratios $r_s^F$ for a weaker disorder. 
However, the use of a dimensionless ratio ignoring $W$ when $W$ is 
large can be misleading. 

\subsection{Divergence of a characteristic length in two dimensions}

\begin{figure}[ht]  
\vskip.2in
\centerline{\epsfxsize=13cm\epsffile{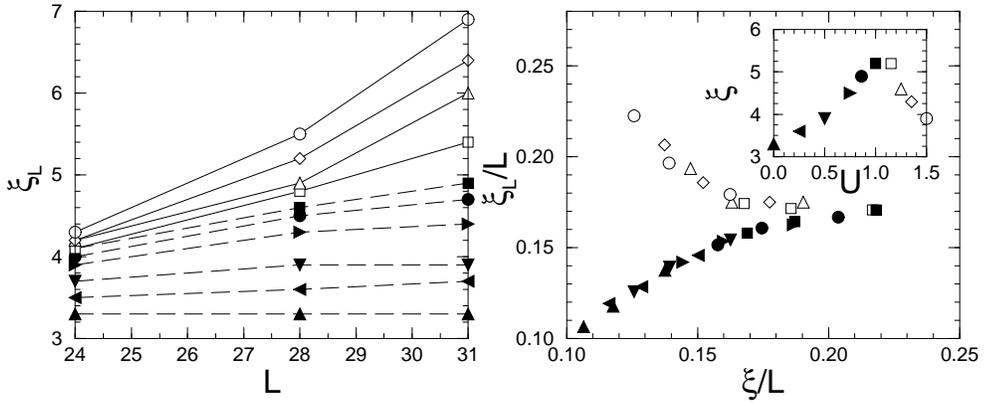}} 
\caption{ 
Right: Characteristic length $\xi_L$ as a function of $L$ for 
$U \leq 1$ (filled symbols): $0$ (triangles up), $0.25$ (triangles left), 
$0.5$ (triangles down), $0.75$ (triangles right), $0.85$ (circles) 
and $1$ (squares) and $U >1$ (empty symbols): $1.15$ (squares), $1.25$ 
(triangles), $1.35$ (diamonds) and $1.5$ (circles).
Left: Ratios $\xi_L/L $ mapped onto the scaling curve $f$ 
as a function of $\xi/L$. The two dimensional scaling length 
$\xi$ is given in the inset.   
}
\label{FIG-scaling} 
\end{figure} 

 From a finite size study \cite{epl1}, one can obtain scaling laws 
consistent with the divergence of a characteristic length of the 
two dimensional system at a first lower threshold $U_c \approx t$, 
as in a second order phase transition. For defining a characteristic 
length of the $N$-body ground state, one considers the change 
$\delta \rho_j$ of the charge density induced by a small change 
$\delta v_i$ of the random potential $v_i$ located at a distance 
$r=|i-j|$.  To improve the statistical convergence, one 
computes the change $\delta \rho (r) = 
\sum_{j_y} \delta \rho_{r,j_y}$ of the charge density on the $L$ 
sites of coordinate $j_x=r$ yielded by the change $v_{0,i_y} 
\rightarrow 1.01 v_{0, i_y}$ for the $L$ random potentials 
of coordinate $i_x=0$. For a Slater determinant made with N occupied  
single particle eigenfunctions ($\psi_\alpha$), first order 
perturbation theory gives:
\begin{equation} 
\delta\rho_j = 2\delta v_i\sum_{\alpha=1}^N\sum_{\beta\neq \alpha}
\frac{\psi_\alpha(i)\psi_\beta(i)\psi_\alpha(j)\psi_\beta(j)}{E_\beta
-E_\alpha}\propto\exp-\frac{2r}{L_1},
\end{equation} 
the index $\beta$ varying over the one-body spectrum. Therefore, 
in the absence of interaction, the change $\delta \rho$ remains 
localized on a scale given by the one-body localization length 
($\xi_L \approx L_1/2$). 

Let us study the dependence of $\xi_L$ on the interaction $U$ for 
an ensemble of $5\times 10^3$ clusters, with $N=3,4,5$ particles in 
square lattices of size $L=24,28,31$ respectively, corresponding to 
a very low filling factor $n_e \approx 5\times 10^{-3}$. 
To have Anderson localization inside these     
sizes we considered a large disorder to hopping ratio  
$W/t=10$. Therefore the low energy tail of the one body spectrum 
is made of impurity states trapped at some site $i$ of exceptionally 
low $v_i$. As we are interested in studying the effect of Coulomb 
repulsion on genuine Anderson localized states we get rid of the 
band tail. Typically we ignore the $L^2/2$ first one-body levels   
(but results do not change, provided that the Fermi level is out 
of the band tail, $\epsilon_F>-4t$).  
From this restricted subset of one-body states we built a basis for 
the $N$-body problem, truncated to the 
$N_H=10^3$ Slater determinants of lowest energy (convergency tests 
are discussed in Ref. \cite{epl1}).
We note that, for $W/t=10$, the one-body localization 
length $L_1\approx 4$ is smaller than the distance between particles 
$d\approx 15$.  

We checked that $|\delta \rho(r)|$  
is reasonably fitted by a log-normal distribution. Therefore it  
makes sense to characterize the typical strength of the fluctuations by
\begin{equation} 
\delta \rho_{\mbox{typ}} (r) = \exp <\ln | \delta \rho (r)| > 
\end{equation} 
and extract the length $\xi_L$ over which the perturbation is effective 
from the exponential decay 
\begin{equation} 
\delta \rho_{\mbox{typ}} (r) \propto  
\exp(-r/\xi_L).
\end{equation}  

The size dependence of $\xi_L$ is presented in Fig. \ref{FIG-scaling} left, 
for increasing Coulomb repulsions. One finds the behavior typical of a 
phase transition: $\xi_L$ converges towards a finite value when $U <U_c$ 
(localized phase), diverges linearly as a function of $L$ at 
$U=U_c\approx 1$ (critical point) and diverges faster 
than linearly when $U>U_c$ (delocalized phase). This is exactly the 
behavior \cite{jlp,kramer} which characterizes the one-body 
Anderson transition in three dimensions. 

In Fig. \ref{FIG-scaling} right we verify a usual scaling ansatz \cite{jlp}  
inspired by the theory of second-order phase transitions: 
\begin{equation} 
\label{ansatz} 
\frac{\xi_L}{L}=f\left(\frac{L}{\xi}\right),  
\end{equation}
where we assume that it is possible to map the characteristic 
length $\xi_L$ at the system size $L$ onto a scaling curve 
$f(L/\xi)$, where $\xi$ is the scaling length characteristic of 
the infinite two dimensional system. 

All the data of Fig. \ref{FIG-scaling} left can be mapped onto the universal 
curve $f$ shown in Fig. \ref{FIG-scaling} right, assuming the scaling 
length $\xi$ given in the inset. When $U<U_c$, this length characterizes 
the localization of the effect of a local perturbation of the substrate 
in the two dimensional thermodynamic limit. Our data are consistent 
with a divergence of $\xi$ at a threshold $U_c \approx t$. 

Very often, additional corrections 
$\propto L^{-\alpha}$ to the scaling ansatz occurs for small sizes. 
We point out that our results can be fitted by a simple linear law $\xi_L =  
0.17 L$ for $U=U_c$. This is a further indication that the simple ansatz 
(\ref{ansatz}) describes scaling for $L \geq 24$, without 
noticeable additional $L^{-\alpha}$ corrections.  

 This tells us that the polarized electron system in a highly 
disordered $2d$ substrate becomes correlated when $U>U_c$ and 
that the effect of a local perturbation becomes delocalized.

\subsection{Size independent distribution of the first excitation}

In Ref. \cite{hfepjb} we used the configuration interaction method,  
discussed in appendix, to study the statistics of the first 
many-body energy spacing. Questions related to the 
convergence of the method when the Hartree-Fock basis is truncated 
are discussed in Ref. \cite{hfepjb}.   

Let us consider the first $N$-body energy levels $E_i$, $(i=0,1,2,...)$
for different sizes $L$, with a large disorder to hopping ratio 
$W/t=15$ imposed to have Anderson localization and Poissonian spectral 
statistics for the one particle levels at $U=0$ when $L\geq 8$. 
This corresponds to a strongly disordered limit where the one particle 
localization length $L_1$ is not only smaller than $L$ and the distance 
$d$ between the particles (as in the previous subsection),
but becomes also of the order of the lattice spacing. 
We considered $N=4,9$, and $16$ particles inside clusters of size $L=8,12$, 
and $16$ respectively. This corresponds to a constant low filling  
factor $n_e=1/16$. We studied an ensemble of $10^4$ disorder configurations.  

\begin{figure}[ht]  
\vskip.2in
\centerline{\epsfxsize=10cm\epsffile{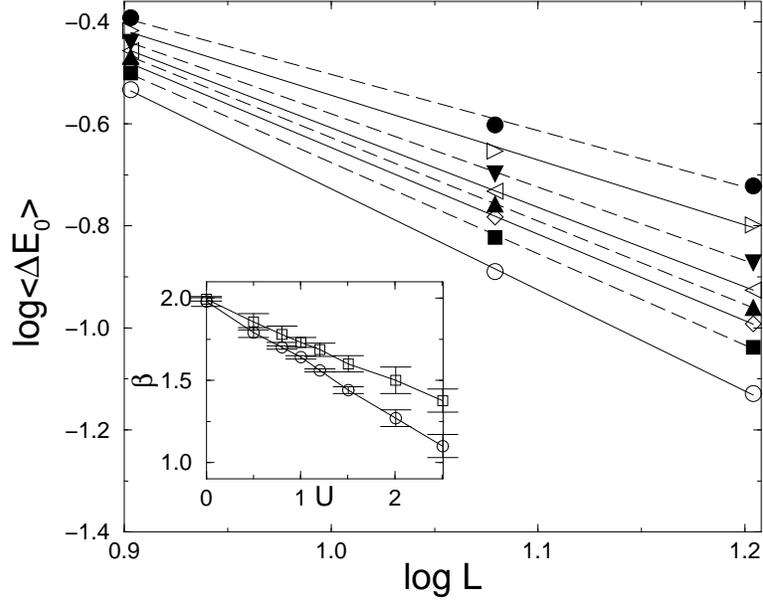}} 
\caption{Size dependence of the average gap 
(first spacing $<\Delta E_0> \propto L^{-\beta (U)}$), 
for $W=15$, filling factor $n_e=1/16$. 
From bottom to top: $U=0,0.5,0.8,1,1.2,1.5,2,2.5$. 
Inset: $\beta (U)$ (circles, characterizing $<\Delta E_0>$ and 
squares, characterizing $<\Delta E_i>$, with an average over $i=1-3$).  
}    
\label{FIG-hf1}
\end{figure} 

The first average spacing $<\Delta E_0>$ is given in Fig. \ref{FIG-hf1}. 
It exhibits a power law decay as $L$ increases, with an exponent $\beta$ 
given in the inset. 
One finds for the first spacing that $\beta$ linearly decreases from 
$d=2$ to $1$ when $U$ increases from $0$ to $2.5$. 
The next mean spacings $<\Delta E_i>=<E_{i+1}-E_i>$ depend more weakly 
on $U$, as shown in Fig. \ref{FIG-hf1}.

\begin{figure}[ht]  
\vskip.2in
\centerline{\epsfxsize=6.5cm\epsffile{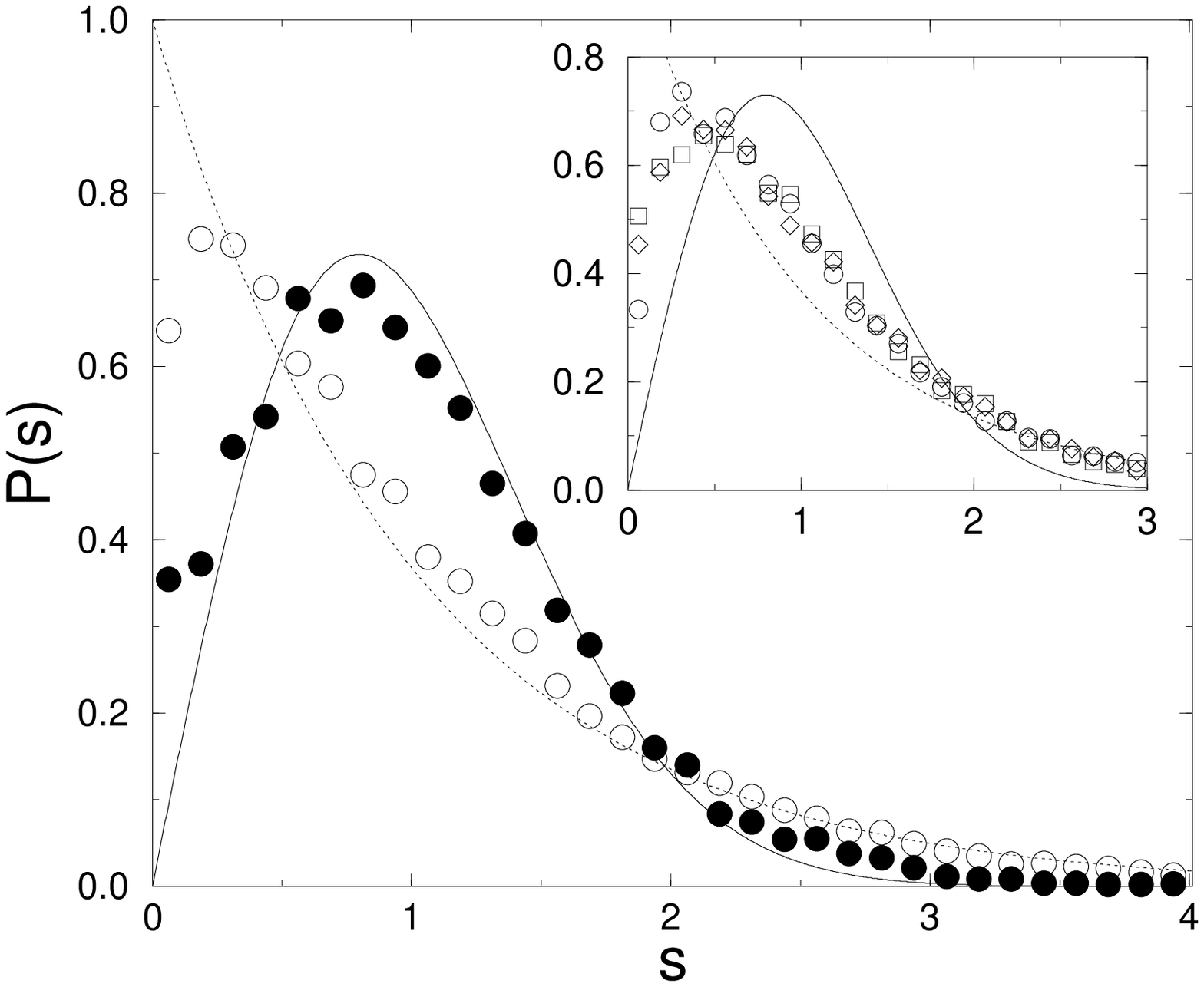}\hfill 
\epsfxsize=6.5cm\epsffile{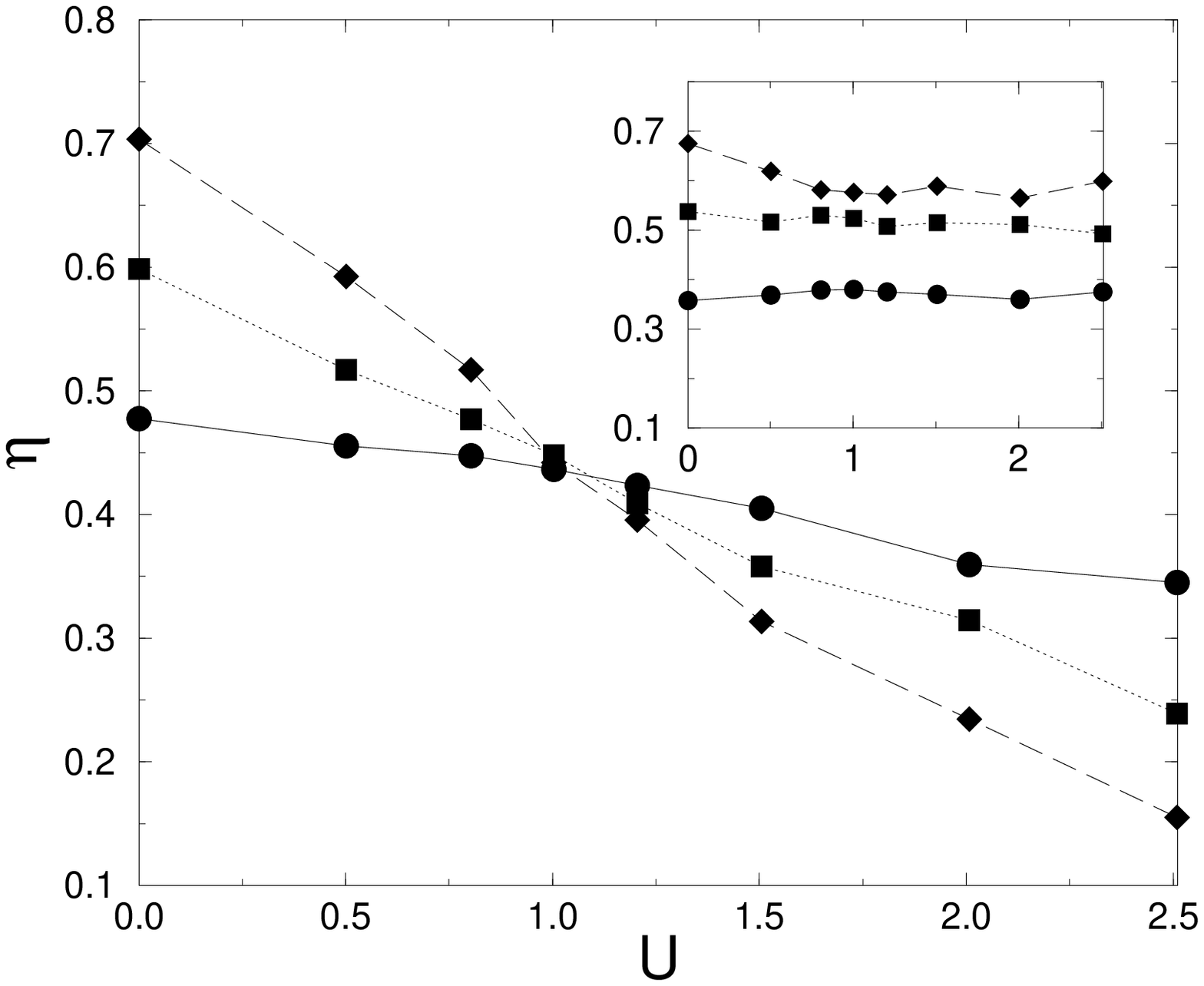}}
\caption{Left: gap distribution $P(s)$ for $U=0$ (empty circles) and 
$U=2.5$ (filled circles) when $N=16$, $L=16$, $W=15$, compared to 
$P_P(s)$ and $P_W(s)$. Inset: size invariant $P(s)$ at 
$U_c\approx 1.1$ ; $L=8$ (circles), $12$ (squares), and $16$ (diamonds). 
Right: parameter $\eta_O(U)$ corresponding to the first spacing $\Delta E_0$ 
at $L=8$ (circles), $12$ (squares), and $16$ 
(diamonds). Inset: $\eta_O(U)$ for the second spacing $\Delta E_1$. 
}    
\label{FIG-hf2}
\end{figure} 

For $U=0$, the distribution of the first spacing $s=\Delta E_0/<\Delta E_0>$
becomes closer and closer to the Poisson distribution $P_P(s)$ 
when $L$ increases, as it should be for an Anderson insulator. For a 
larger $U$, the distribution seems to become close to the Wigner surmise 
$P_W(s)$ characteristic of level repulsion 
in Random Matrix Theory, as shown in Fig. \ref{FIG-hf2} left, for $U = 2.5$ 
and $L=16$. To study how this $P(s)$ goes from Poisson to a Wigner-like 
distribution when $U$ increases, we use the spectral parameter $\eta_O$ 
which decreases from $1$ to $0$ when $P(s)$ goes from Poisson to Wigner. 
In Fig. \ref{FIG-hf2} right, one can see that the three curves $\eta_O (U)$ 
characterizing the first spacing for $L=8, 12, 16$ intersect at 
a critical value $U_c/t \approx 1.1$. Our data suggest that 
for $U < U_c$ the distribution tends to Poisson in the thermodynamic 
limit, while for $U>U_c$ it tends to a Wigner-like behavior. 
At the threshold $U_c$, there is a size-independent intermediate 
distribution shown in the inset of Fig. \ref{FIG-hf2} left, exhibiting level 
repulsion at small $s$ followed by a $\exp( -a s)$ decay at large $s$,  
with $a \approx 1.52$. Such a size independent distribution 
is known for characterizing a mobility edge in an one body spectrum. 
This Poisson-Wigner transition characterizes 
only the first spacing, the distributions of the next spacings being 
quite different. The inset of Fig. \ref{FIG-hf2} right does not show any  
intersection for the parameter $\eta$ calculated for the second 
spacing. 

The observed transition, and the difference between the first spacing and 
the following ones is mainly an effect of the H-F mean field. 
For the first spacings, the curves $\eta_O$ calculated with the HF data are 
qualitatively similar. At the mean field level the low energy excitations 
are particle-hole excitations starting from the ground state. 
The energy spacing between the first and the second excited states is 
given by the difference of two particle-hole excitations and a 
Poisson distribution follows if the low energy particle-hole 
excitations are uncorrelated. 

We note that the energy of an electron-hole pair is given by 
$\epsilon_j-\epsilon_i-U/r_{ij}$ and the classical argument for the 
existence of a gap in the single particle density of states does not 
apply for the many-body spectrum\cite{ES}. 
Therefore the observed opening of a gap for the 
first energy excitation is a remarkable phenomenon beyond the 
predictions of the classical Coulomb gap model.  

\subsection{Change of the inverse compressibility}

In Ref. \cite{hfepjb} we studied also the inverse compressibility  
\begin{equation} 
\label{invcomp} 
\Delta_2(N)=E_0(N+1)-2E_0(N)+E_0(N-1),   
\end{equation} 
i.e. the discretized second derivative of the ground state energy 
$E_0$ with respect to the number $N$ of particles. 

We consider $N=4,9,16$ particles on square lattices of size 
$L=6,9,12$ respectively, corresponding to a constant filling 
factor $n_e=1/9$. Here we focus on the strongly localized 
regime with disorder strength $W=15$.  

The following inverse compressibility data are obtained with the 
configuration interaction method. 
We have checked that the residual interaction does not change 
qualitatively Hartree-Fock results. 
Indeed HF approximation gives in this weak coupling regime a 
good estimate of the ground state 
energy (see Ref. \cite{hfepjb}) and the inverse compressibility 
is a physical observable which only depends on the ground state 
energies at different number of particles. 
Neither a precise knowledge of the ground state wavefuction (as in 
the calculation of persistent currents) nor excited states energies 
(as in studies of spectral statistics) are required. 

\begin{figure}[ht]  
\vskip.2in
\centerline{\epsfxsize=6.5cm\epsffile{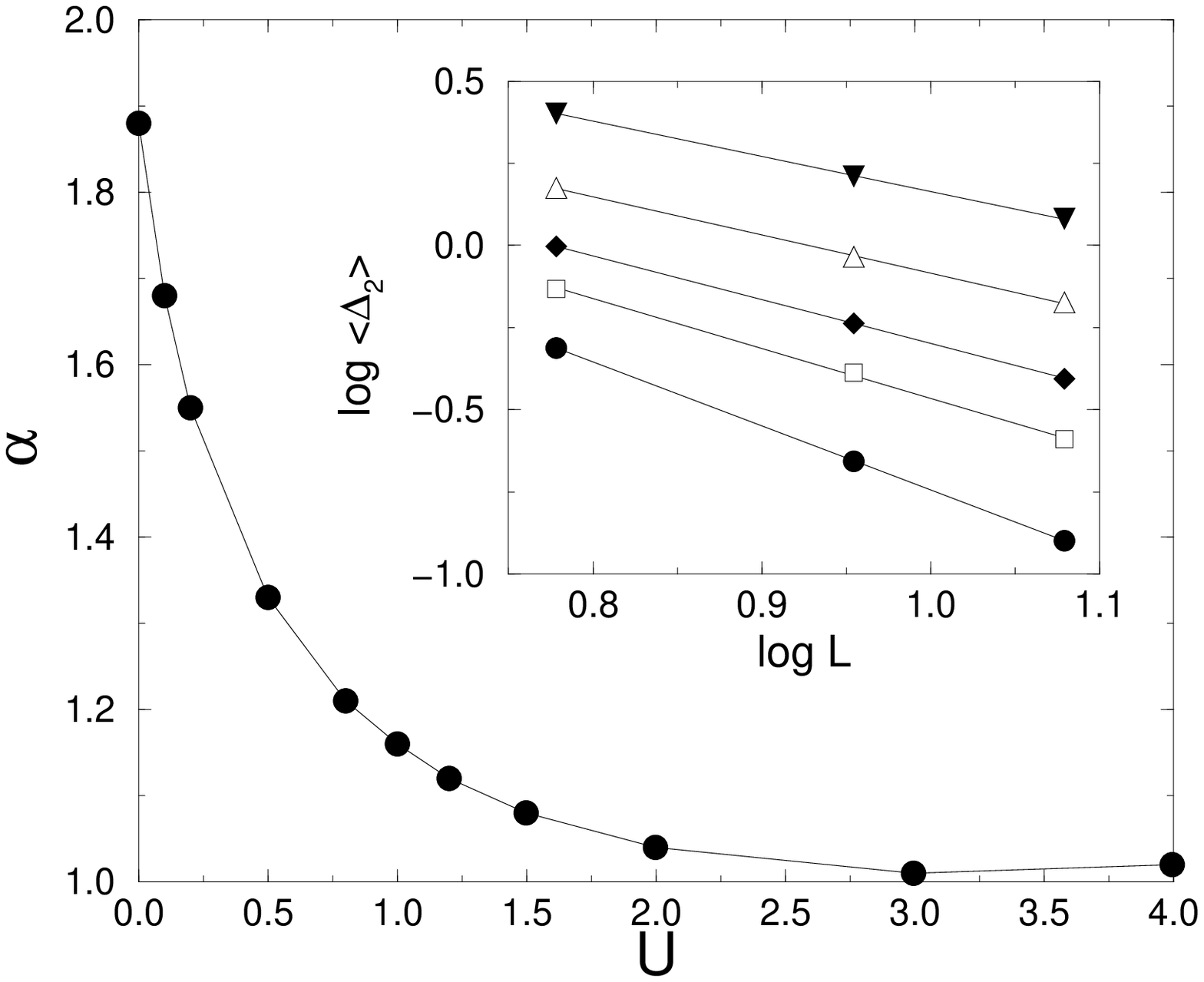}\hfill 
\epsfxsize=6.5cm\epsffile{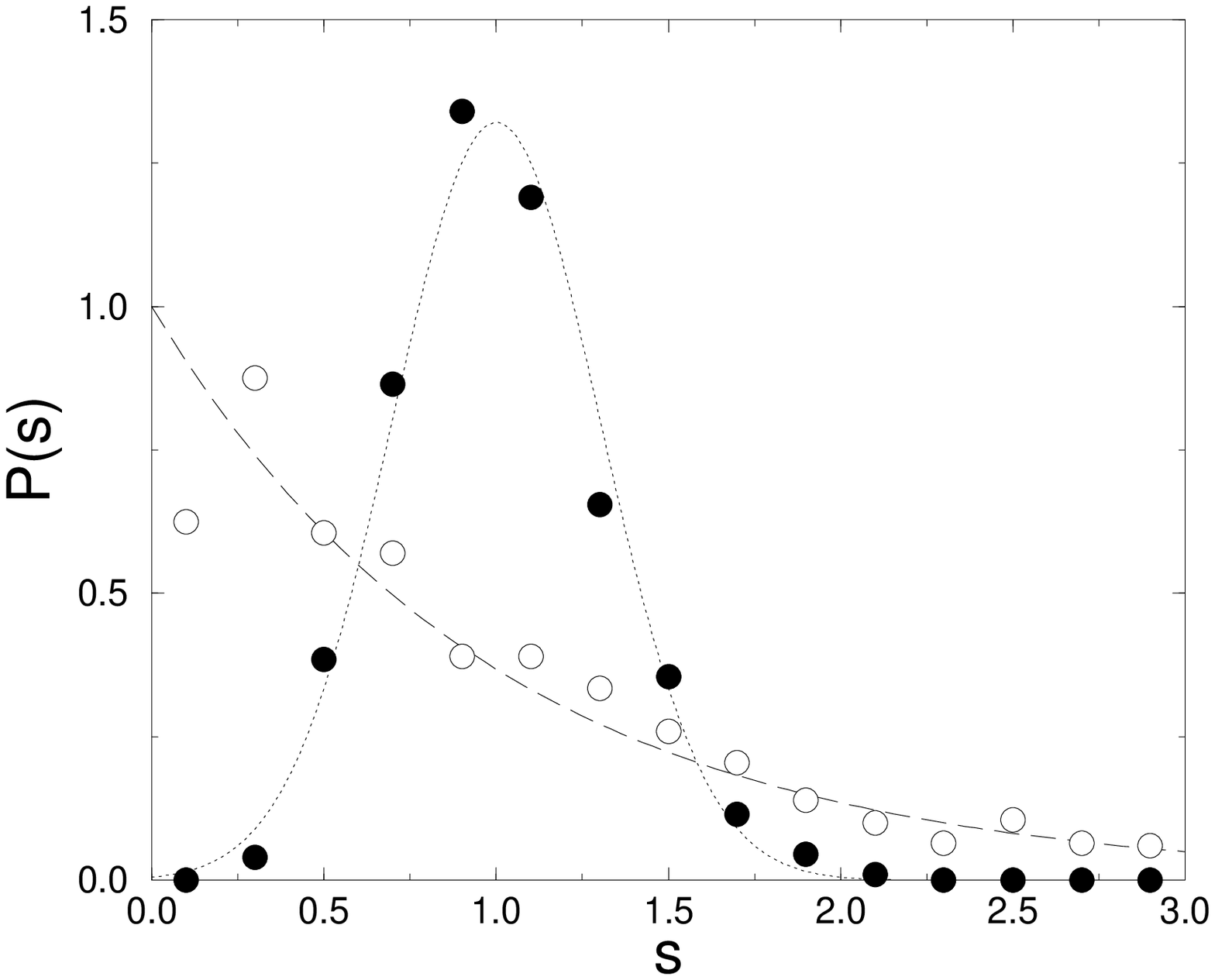}}
\caption{ 
Left: Inset: size dependence of the average inverse compressibility 
$<\Delta_2>$, for $W=15$, filling factor $n_e=1/9$, $U=0$ (circles), 
$0.5$ (squares), $1$ (diamonds), $2$ (triangles up), and $4$ (triangles down). 
Straight lines are power law fits $<\Delta_2(U)>\propto
L^{-\alpha(U)}$. Main figure: exponent $\alpha(U)$. 
Right: distribution of the normalized inverse compressibilities 
for $N=16$, $L=12$, $W=15$, $U=0$ 
(empty circles) and $U=3$ (filled circles), fitted by a Gaussian 
of standard deviation $\sigma=0.30$ (dotted line). Dashed 
line gives Poisson distribution.  
Disorder average is over $10^3$ configurations. 
}
\label{FIG-hf3}    
\end{figure} 
 
Fig. \ref{FIG-hf3} left shows the $L$-dependence of the average inverse 
compressibility, which is well fitted by the power law  
$<\Delta_2(U)>\propto L^{-\alpha(U)}$, with $\alpha(U)$ 
going from $2$ to $1$ when $U$ goes from $0$ to $3$ approximately. 
The value $\alpha=2$ is expected without interaction 
($<\Delta_2>=<\Delta>\propto 1/L^2$, with $<\Delta>$ single 
particle mean level spacing). 
The exponent $\alpha=1$, in this strongly localized regime, 
can be related to the Coulomb gap in the single particle density 
of states \cite{ES}: According to Koopmans' theorem (see 
Refs. \cite{Levit,Walker,Berkovits} for a 
thorough discussion about the limits of validity of the Koopmans' 
theorem for disordered quantum dots), one assumes 
that all the other charges are not reorganized by the 
addition of an extra charge, 
\begin{equation} 
\Delta_2\approx \epsilon_{N+1}-\epsilon_N\propto \frac{1}{L}  
\end{equation}
due to the Coulomb gap, where $\epsilon_k$ is the energy of the 
$k$-th HF orbital at a fixed number $N$ of particles.  
 
The distribution of inverse compressibilities, 
for $L=12$ and $W=15$ is shown in Fig. \ref{FIG-hf3} right. 
At $U=0$, due to Anderson localization, $\Delta_2$ distributions are 
close to the Poisson distribution 
(deviations from the Poisson distribution at 
small $s$ values are due to the finite system size). 
On the contrary, the distribution at $U=3$ shows a Gaussian shape. 
This can be understood within the Koopmans' theorem, since the 
HF energies are given by 
\begin{equation}
\epsilon_k=\langle\psi_k| H_1 |\psi_k\rangle +\sum_{\alpha=1}^N 
\left(Q_{\alpha k}^{\alpha k}-Q_{\alpha k}^{k \alpha}\right),      
\end{equation} 
with $H_1$ one-body part of the spinless Hamiltonian 
and $Q_{\alpha\beta}^{\gamma\delta}$ interaction matrix elements 
given by Eq. (\ref{matel}) in appendix C. 
Due to the small correlations of eigenfunctions in a random 
potential, one can reasonably invoke the central limit theorem
(see Ref. \cite{Berkovits} for a more detailed discussion).  
We point out that a Gaussian-like distribution 
of inverse compressibilities has been observed in quantum 
dots experiments in the Coulomb blockade regime 
\cite{Sivan,Marcus,Simmel}.  
This implies that inverse compressibility fluctuations are 
dominated by interaction effects instead of single particle 
fluctuations. 

We note that the Gaussian-like shape of 
the inverse compressibility distribution becomes more 
asymmetric when the disorder strength is increased. 
This asymmetry is due to the Coulomb gap and has 
been discussed in the classical limit ($t=0$) in
Ref. \cite{Shklovskii}.  

A simple model for the addition spectra of quantum dots 
has been recently proposed by B. Shapiro \cite{shapiro}. 
This model explains the change of the inverse compressibility 
distribution when $r_s$ increases assuming that the system can be 
decomposed into a stable N-electron Wigner lattice plus a lattice 
of interstitial sites where the added particle can move and for 
which a usual single particle description is assumed. This 
suggests possible relations between the problem of adding an 
extra particle to a Wigner solid and its melting when $r_s$ 
decreases. 

\section{Conclusion}

 We have numerically revisited the quantum-classical crossover 
from the weak coupling Fermi limit towards the strong coupling 
Wigner limit, using small lattice models. In the continuous limit, 
the assumed picture \cite{tanatar-ceperley} is relatively simple: a 
liquid-solid first order transition for $r_s \approx 37$. For 
a clean lattice model at a filling factor $1/9$, our results 
raise the question of the possible existence of an intermediate 
liquid-solid phase as first proposed by Andreev and Lifshitz. 
This may also raise questions about the differences between the 
continuous limit and lattice models. When the spin degrees of 
freedom are included, our lattice model is a Hubbard 
model with additional long range repulsions far from half 
filling. The physics of such models is not very well known, but 
at least suspected to be very rich and complex, as discussed in 
many works since the discovery of high $T_c$ superconductors. 
The role of the disorder is another source of complexity. We have 
divided this chapter in two main parts, the first where the 
disorder is weak, the second where the disorder is strong and 
makes the quantum kinetic effects less relevant. This raises also 
the question of the difference between the quantum Wigner glass 
and the classical Coulomb glass. For large disorder, a finite size 
scaling analysis leads us to conclude that the weak coupling 
Fermi phase is delimited by a second order quantum phase transition, 
when $r_s$ increases. 

Eventually, our first motivation was the experimental discovery of 
a possible new metal for intermediate values of $r_s$. 
Our numerical studies have not directly addressed the transport 
properties for intermediate $r_s$. One cannot claim that the 
observed intermediate regime gives a new metal in the thermodynamic 
limit. Nevertheless, if transport is mainly due to the presence of 
delocalized defects in a floppy and pinned electron solid for 
intermediate $r_s$ , it might be interesting to extend the FLT 
theory \`a la Landau proposed in Ref. \cite{DKL} for a clean system 
to the case where one has also elastic scattering by impurities. 
It will tell us if a modified FLT adapted to a very special Fermi 
system with a non conventional Fermi surface, still gives rise to 
usual quantum interferences leading to weak localization and eventually 
to Anderson localization for the conjectured gapless excitations 
of a floppy solid. This could help to determine whether one has 
a true new metal for intermediate $r_s$ and weak disorder, or only an 
``apparent'' metallic behavior which should disappear at the true 
$T \rightarrow 0$ limit. This might help to explain why in measurements 
down to $5 mK$ in a GaAs hetrostructure \cite{bell-labs}, the weak 
localization correction remains less than a few percent of 
the value predicted for a standard disordered $2d$ Fermi liquid.

\chapappendix{Hartree-Fock approximation}

This is the usual mean field approximation when one has electron-electron 
interactions, which we shortly review for the case of spinless fermions. 
The HF ground state (GS) is the Slater Determinant 
(SD) which minimizes the GS energy expectation value. In the 
HF approximation, one reduces the two-body part of the 
total Hamiltonian to an effective 
single particle Hamiltonian 
\cite{Kato,Poilblanc,Schreiber}
\begin{equation} 
\label{hfhamiltonian} 
\begin{array}{c}
\displaystyle{ 
U (\sum_{i\neq j} \frac{1}{r_{ij}} n_i\langle n_j \rangle
- \sum_{i\neq j} \frac{1}{r_{ij}} c^{\dagger}_i c_j  
\langle c^{\dagger}_j c_i \rangle }),  
\label{HF}
\end{array} 
\end{equation} 
where $\langle ... \rangle$ stands for the expectation value with 
respect to the HF ground state, which has to be determined 
self-consistently. 
The first (Hartree) term describes the interaction of any electron 
with the charge distribution set up by all the other electrons, 
while the latter (Fock) term is a nonlocal exchange potential. 
The Hartree term comes out as an extra on--site disorder potential, 
while the Fock term introduces extra hopping amplitudes.

The main advantage of the Hartree-Fock approximation is that it 
reproduces the single--particle density of states 
$g(E)$ \cite{Schreiber}, particularly the Coulomb gap 
near the Fermi energy $E_F$: in the two--dimensional case,  
$g(E)\propto |E-E_F|$ \cite{ES}. 
The physical argument underlying this relation is that an empty 
site $j$ and an occupied site $i$ with a difference in energy 
$\epsilon_j-\epsilon_i$ 
smaller than $\delta$ must be at a distance larger than 
$U/\delta$ as the change $\Delta E_{ij}=\epsilon_j-
\epsilon_i-U/r_{ij}$ of the system energy must be positive 
when we consider an excitation starting from the ground state. 
However, the HF approximation gives a Coulomb gap also in the 
delocalized regime at small disorder (single particle localization 
length larger than the system size), where complicated many-body 
effects beyond HF approximation should screen Coulomb interaction.  

\chapappendix{Stoner instability}

Ferromagnetic instabilities come from the interplay 
between Coulomb repulsion and the Pauli principle. 
In the Pauli picture, electrons populate the orbital states 
of a system, such as a quantum dot or a metallic grain, in a 
sequence of spin up - spin down electrons. 
The resulting minimum spin state minimizes the kinetic 
energy: it costs energy to flip a spin since it must be 
promoted to a higher energy level. On the other hand, 
the maximum spin state requires a maximally antisymmetric 
coordinate wavefunction, thus reducing the effect of 
Coulomb repulsion. This is at the basis of Hund's rule 
for atoms: electrons occupy orbitals in an open shell so 
as to maximize their total spin.  

Due to the locality of the Pauli principle, ferromagnetic 
instabilities can be studied within the Hubbard Hamiltonian, 
\begin{equation} 
\label{hubbard}
H=-t\sum_{<i,j>\sigma} c^{\dagger}_{i\sigma} c_{j\sigma} +  
\sum_{i\sigma} v_i n_{i\sigma}  + 
U \sum_{i} n_{i\uparrow} n_{i\downarrow}.
\end{equation} 

In the HF approximation the interaction part of the Hubbard 
Hamiltonian (\ref{hubbard}) is reduced to 
\begin{equation} 
U\sum_{i\sigma} n_{i\sigma} \langle n_{i -\sigma} \rangle.  
\end{equation}   
The HF energies are given by 
\begin{equation} 
\epsilon_{\alpha\sigma}=\epsilon_\alpha^0 + 
U \sum_{i\alpha\beta} |\psi_\alpha (i)|^2 |\psi_\beta (i)|^2 
\langle n_{\beta -\sigma} \rangle,   
\end{equation}   
with $\epsilon_\alpha^0$ one-body eigenenergies and 
$n_{\beta -\sigma}=d_{\beta -\sigma}^\dagger d_{\beta -\sigma}$ 
occupation numbers for the HF orbitals. The total ground state HF  
energy reads 
\begin{equation} 
E_0=\sum_{\alpha\sigma}\epsilon_\alpha^0 
\langle n_{\alpha\sigma} \rangle  + 
U \sum_{i\alpha\beta} |\psi_\alpha (i)|^2 |\psi_\beta (i)|^2 
\langle n_{\alpha \uparrow} \rangle   
\langle n_{\beta \downarrow} \rangle.    
\end{equation}   

In order to study the stability of the nonmagnetic solution, 
\begin{equation} 
\langle n_{\alpha\sigma} \rangle=
\left\{
\begin{array}{l}
1 \,\,\,\, \hbox{if} \,\,\,\, 
\epsilon_{\alpha\sigma} < \epsilon_F \\  
0 \,\,\,\, \hbox{if} \,\,\,\, 
\epsilon_{\alpha\sigma} > \epsilon_F,  
\end{array} 
\right.
\end{equation} 
with $\epsilon_F$ Fermi level, we take a layer (of thickness 
$\delta \epsilon$) of electrons with spin down below the Fermi 
level to put them in states with spin up \cite{foner}.   
This changes the spin state of $\delta n_s = \rho(\epsilon_F)
\delta\epsilon$ electrons per unit volume, where  
$\rho (\epsilon_F)$ is the density of states per 
volume per spin. 
The change of the one-body energy density is given by 
\begin{equation} 
\delta w_0 = \rho (\epsilon_F) (\delta\epsilon)^2. 
\end{equation} 
In the clean case, the change of the interaction energy 
(per volume) is   
\begin{equation} 
\delta w_{\hbox{int}}=U \left( \frac{n_s}{2} + \rho \delta\epsilon \right)  
\left( \frac{n_s}{2} - \rho \delta\epsilon \right)  
- \frac{U n_s^2}{4} = - U \rho ^2 (\epsilon _F) (\delta \epsilon)^2. 
\end{equation} 
Therefore the nonmagnetic state becomes unstable when 
$\delta w_0 +\delta w_{\hbox{int}}<0$, that is 
\begin{equation} 
\label{stoner} 
U \rho (\epsilon_F) >1,  
\end{equation}  
which gives the Stoner criterion for ferromagnetic instability. 

In the diffusive regime, the effective interaction strength is enhanced 
by the presence of disorder, leading to ferromagnetic instabilities 
already below the Stoner threshold, as pointed out in Ref. 
\cite{kamenev}.  

We also note that Stoner criterion predicts a ferromagnetic 
ground state for the Hubbard model even at finite temperatures 
in one and in two dimensions, thus violating the Mermin and Wagner's 
theorem. Therefore, when the Stoner criterion (\ref{stoner}) 
is satisfied, we can only conclude that the nonmagnetic 
ground state is unstable. 

 The problem of a possible magnetization of the ground state is not only 
discussed in a dilute 2DEG, but is also central in the studies of 
mesoscopic quantum dots since their Ohmic resistances are measured as a 
function of a gate voltage in the Coulomb blockade regime. The possibility 
of a spontaneous magnetization $S$ of their ground state due to 
electron-electron interactions has been proposed to explain the 
observed conductance peak spacing distributions.

\chapappendix{Configuration interaction method}

Even though the approximations involved in the HF method are 
uncontrolled, the mean field HF results can be improved using 
a numerical method \cite{CIphys1,CIphys2,hfepjb} familiar in quantum 
chemistry as the configuration interaction method (CIM) 
\cite{CIchem}. Once a complete orthonormal basis of HF orbitals 
has been calculated, 
\begin{equation} 
\label{orbitals} 
H_{HF}(|\psi_1\rangle,...,|\psi_N\rangle)  
|\psi_{\alpha}\rangle = \epsilon_{\alpha}|\psi_{\alpha}\rangle,    
\end{equation} 
with $ \alpha=1,2,\ldots,L^2$, it is possible to build up a Slater 
determinants' basis for the many-body problem which can be truncated to 
the $N_H$ first Slater determinants, ordered by increasing energies. 
The two-body Hamiltonian can be written as 
\begin{equation}
H_{\rm int}=\frac{1}{2} \sum_{\alpha,\beta,\gamma,\delta}  
Q_{\alpha\beta}^{\gamma\delta} d_{\alpha}^{\dagger} d_{\beta}^{\dagger} 
d_{\delta}d_{\gamma} ,
\label{Hint}
\end{equation} 
with 
\begin{equation} 
Q_{\alpha\beta}^{\gamma\delta} = 
U\sum_{i\neq j} \frac{ \psi_\alpha(i) \psi_\beta(j) \psi_\gamma(i) 
\psi_\delta (j)}{r_{ij}}  
\label{matel}
\end{equation}
and $d^{\dagger}_{\alpha}=\sum_j \psi_{\alpha} (j) c^{\dagger}_j $. 
One gets the residual interaction subtracting Eq. \ref{HF} from  
Eq. \ref{Hint}.

\begin{acknowledgments}
The authors wish to acknowledge the participation of D. Shepelyansky 
in one of the reviewed works (Ref. \cite{hfepjb}) and to thank 
B. Spivak for discussions about the work of Andreev and Lifshitz.
\end{acknowledgments}

Present addresses:

G. Benenti: Universit\`a degli Studi dell'Insubria
and Istituto Nazionale per la Fisica della Materia,
via Valleggio 11, 22100 Como, Italy. 

G. Katomeris: Department of Physics, University of 
Ioannina, 45 110, Greece. 

X. Waintal: Laboratory of Atomic and Solid State Physics
530 Clark Hall, Cornell University, Ithaca, NY 14853-2501,  USA.

\end{document}